\documentclass{aa}
\usepackage{psfig}

\def\la{\;
\raise0.3ex\hbox{$<$\kern-0.75em\raise-1.1ex\hbox{$\sim$}}\; }
\def\ga{\;
\raise0.3ex\hbox{$>$\kern-0.75em\raise-1.1ex\hbox{$\sim$}}\; }

\newcommand{\zabs}{$z_{\rm abs}\,$}
\newcommand{\zem}{$z_{\rm em}\,$}

\newcommand{\kms}{km~s$^{-1}\,$}
\newcommand{\cm}{cm$^{-2}\,$}
\newcommand{\cmm}{cm$^{-3}\,$}

\begin{document}

\title{Quasar spectral energy distribution in EUV restored 
from associated absorbers:
indications to the \ion{He}{ii} opacity of the quasar accretion disk 
wind\thanks{Based on observations obtained at the VLT Kueyen telescope 
(ESO, Paranal, Chile), and at the HST.} 
}

\author{
S. A. Levshakov\inst{1,3}
\thanks{On leave from the Ioffe Physico-Technical Institute, 
St. Petersburg, Russia}
\and
I. I. Agafonova\inst{2,3}
\and
D. Reimers\inst{1}
\and
J. L. Hou\inst{3}
\and
P. Molaro\inst{4}
}

\offprints{S.~A.~Levshakov
\protect \\lev@astro.ioffe.rssi.ru}

\institute{
Hamburger Sternwarte, Universit\"at Hamburg,
Gojenbergsweg 112, D-21029 Hamburg, Germany
\and
Ioffe Physico-Technical Institute, 
Polytekhnicheskaya Str. 26, 194021 St.~Petersburg, Russia 
\and
Shanghai Astronomical Observatory, 80 Nandan Road, Shanghai 200030, P.R. China
\and
Osservatorio Astronomico di Trieste, Via G. B. Tiepolo 11,
34131 Trieste, Italy
}

\date{Received 00  / Accepted 00 }

\abstract{}
{To reconstruct the spectral shape of the quasar ionizing radiation
in the extreme-UV range (1 Ryd $\leq E < 10$ Ryd) from
the analysis of narrow absorption lines (NAL) of
the associated  systems.
}
{Computational technique for inverse spectroscopic problems~-- Monte Carlo
Inversion augmented by procedure of the spectral shape recovering and modified
to account for the incomplete coverage of the light source. 
}
{The ionizing spectra responsible for the ionization structure 
of the NAL systems require an intensity depression 
at $E > 4$ Ryd which is attributed to the \ion{He}{ii} Lyman continuum
opacity
($\tau^{\rm He\,{\scriptscriptstyle II}}_c \sim$1). 
A most likely source of this opacity is a quasar 
accretion disk wind.
The corresponding column density of \ion{H}{i} in the wind is 
estimated as a few times $10^{16}$ \cm. 
This amount of neutral hydrogen should cause
a weak continuum depression at $\lambda \la 912$ \AA\ (rest-frame), 
and a broad and shallow absorption in \ion{H}{i}
Ly-$\alpha$. 
If metallicity of the wind is high enough,
other resonance lines of \ion{O}{vi}, \ion{Ne}{vi}~- \ion{Ne}{viii}, etc.
are expected.  In the analyzed QSO spectra we do observe
broad (stretching over 1000s \kms)
and shallow ($\tau \ll 1$) absorption troughs 
of \ion{H}{i} Ly-$\alpha$ and \ion{O}{vi} $\lambda\lambda1031, 1037$   
as well as continuum depressions at $\lambda \la 912$ \AA\ 
which correspond to $N$(\ion{H}{i})~$\sim$$5\times10^{16}$ \cm.
Observational data available in both the UV and X-ray ranges suggest
that at least $\sim$50\% of the quasar radiation
passes through the gas opaque in the \ion{He}{ii} Lyman continuum. 
This means that the outcoming ionizing spectrum should have a pronounced 
intensity break at $E > 4$ Ryd 
with a depth of this break
depending on the angle to the rotational axis of the accretion disk 
(the larger the angle the deeper the break). 
The QSO spectra with a discontinuity at 4 Ryd
can influence the rate of the \ion{He}{ii} reionization 
in the intergalactic medium and partly explain inhomogeneous (patchy)
ionization structure of the intergalactic \ion{He}{ii}  
observed at $z \sim 3$.
}
{}

\keywords{Cosmology: observations --
Line: formation -- Line: profiles -- 
Quasars: absorption lines 
} 

\authorrunning{S. A. Levshakov et al.}
\titlerunning{\ion{He}{ii} opacity of the quasar accretion disk wind}
\maketitle

\section{Introduction}

Spectral energy distribution (SED) of 
the quasar/AGN outcoming radiation in the extreme-UV 
range (EUV, $\lambda < 912$ \AA) is
important for the interpretation of both emission and absorption spectra 
since this range contains the ionization
thresholds of usually observed ions  \ion{C}{ii}-\ion{C}{iv}, 
\ion{N}{ii}-\ion{N}{v},
\ion{O}{i}-\ion{O}{vi}, \ion{Si}{ii}-\ion{Si}{iv}. 
SED is accessible for direct measurements in the FUV-EUV ranges (rest-frame) 
for wavelengths $\lambda > 300$ \AA\  
(e.g., Zheng et al. 1997; Telfer et al. 2002; Scott et al. 2004)  and   
at shorter wavelengths only in
the X-ray range for $\lambda < 40$ \AA\ 
(e.g., Piconcelli et al. 2005; Brocksopp et al. 2006; Costantini et al. 2007). 
In the intermediate range (40 \AA\ $< \lambda < 300$ \AA), direct
measurements of the outcoming radiation of QSOs
with redshifts $z < 2$ are prevented by the Galactic absorption,
and for QSOs with $z > 2$ are, in principle, impossible due to the
intergalactic gas opacity in \ion{He}{ii} Ly-$\alpha$. 

Usually the gap in the SED between
$\lambda = 912$ \AA\ and soft X-ray range is  approximated
by a simple power law, $F_\nu \propto \nu^{-\alpha}$. 
However, this part of the QSO continuum radiation 
is responsible for the ionization state of the so-called narrow absorption line 
(NAL) systems~-- 
metal systems arising in the gas located close to the
quasar host galaxy~-- and can be reconstructed from their analysis. 

The computational procedure aimed at restoring the spectral shape
of the underlying ionizing radiation from ions observed in optically
thin metal absorption-line systems is described in 
Agafonova et al. (2005, 2007). 
Applied to the analysis of NAL systems identified in the spectrum of
the quasar \object{HE 0141--3932} with \zem = 1.8 (Reimers et al. 2005)
this procedure resulted in a SED which  
could not be described with a single power law index $\alpha$:
the EUV spectrum shows a step-like structure 
with a sharp intensity break
at $\lambda \simeq 240$ \AA\ ($E \simeq$ 3.8 Ryd, shifted by $\sim$12 \AA,
or $\sim$16000 \kms\ redward from the \ion{He}{ii} Ly-$\alpha$
edge 228 \AA, or 4 Ryd)  
and a much more slower intensity decay in the range 4 Ryd\ $< E < 10$ Ryd.
Being attributed to the Lyman continuum absorption in \ion{He}{ii},
the depth of the break gives the \ion{He}{ii} column density
$N$(\ion{He}{ii})~= $0.7\times10^{18}$ \cm. 
Similar spectra with breaks at $\sim$4 Ryd corresponding to
$N$(\ion{He}{ii})~= $(0.7-1.4)\times10^{18}$ \cm were reconstructed
from the intergalactic absorption-line systems with \zabs~$\la 1.8$
(Fig.~20 and Table~4 in Agafonova et al. 2007).
Since there are indications that a large part of the QSO/AGN emitted
radiation remains at this redshift unprocessed by the IGM, we argued
that these spectra may represent the energy distribution of the
outgoing quasar radiation.  
In the present paper, we continue the study of the intrinsic quasar
SED using four associated (i.e. physically related to QSO) 
NAL systems identified in spectra
of high-redshift ($2 <$ \zem $< 3$) QSOs.

Concerning the selection of the NAL systems considered in the paper
the following is to be noted.
The metal NAL systems are in general
not rare~-- they are observed in 30\%-50\% of 
quasar spectra (Misawa et al. 2007).  
However, for most of them the observed wavelength coverage and the
ionization conditions in the absorbing gas result in the fact that at best only 
three highly ionized doublets
\ion{C}{iv}$\lambda\lambda 1548, 1551$ \AA, 
\ion{N}{v}$\lambda\lambda 1238, 1242$ \AA, and 
\ion{O}{vi}$\lambda\lambda 1031, 1037$ \AA\ are available for the study.
The column densities of these absorption lines can be reproduced
with a wide variety of the ionizing spectra and, thus, do not allow to distinguish
between specific SEDs.  
Another aggravating factor while working with NAL systems is
that the estimation
of the ion column densities is often hampered by effects like partial
coverage of the background source of the continuum radiation
and/or line blending. 

This explains the selection criteria which governed the search 
for the appropriate associated systems: 
(1) the presence of many lines of different ions~-- to obtain reasonable
restrictions on the SED,  
(2) clear line profiles, and 
(3) the possibility to determine the covering factor with a sufficiently high
accuracy.
These requirements significantly restricted the number of suitable systems, 
but the loss in the quantity was compensated by the increased 
quality of the data which allowed us to clarify some additional
aspects of the physical properties of the NAL systems.

The paper is organized as follows. 
In Sect.~2 we reconstruct SED of the intrinsic
quasar EUV radiation from the 
analysis of quasar-related metal absorption-line systems 
identified in spectra of four QSOs with redshifts \zem = 2.2--2.9. 
The results obtained are discussed in Sect.~3, and summarized in Sect.~4.

\begin{figure*}[t]
\vspace{0.0cm}
\hspace{-0.2cm}\psfig{figure=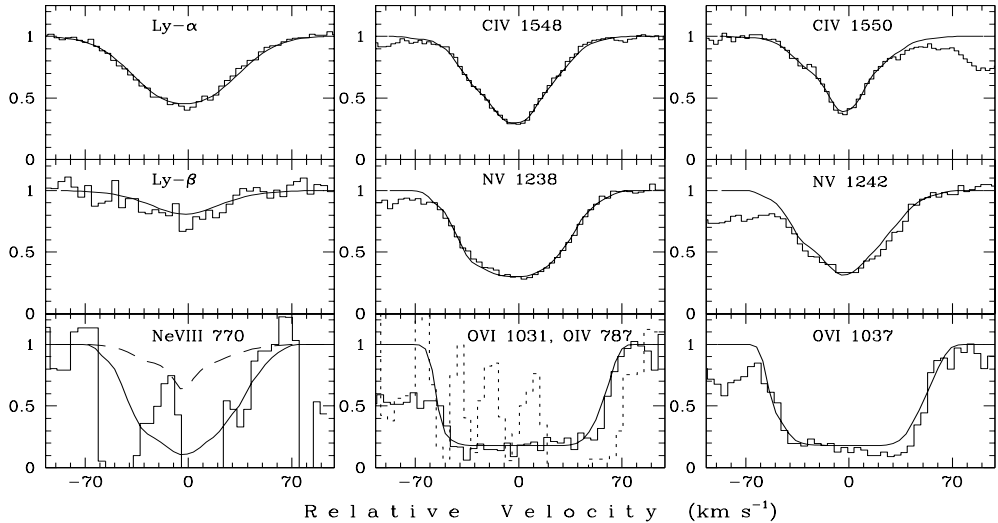,height=16cm,width=18cm}
\vspace{-8.5cm}
\caption[]{
Hydrogen and metal absorption lines from the \zabs = 2.198 system
towards \object{J 2233--606} (solid-line histograms).
The vertical axis is normalized intensities.
The zero radial velocity is fixed at $z = 2.1982$.
Synthetic profiles are plotted by the smooth curves. 
In panel \ion{Ne}{viii}, the dashed curve corresponds to the profile
predicted for the power law ionizing spectrum $F_\nu \propto \nu^{-1.5}$,
and the smooth curve~-- to the profile predicted for the SED with a break
at $\sim$4 Ryd (Fig.~\ref{fg_3}).
The observed profile of \ion{O}{iv}$\lambda 787$ \AA\ (dotted-line histogram)
overplotted on \ion{O}{vi} $\lambda 1031$
is blended with Ly-$\gamma$ from the \zabs = 1.59 system
in the range $\sim$20-70 \kms.
}
\label{fg_1}
\end{figure*}

\section{Analysis of individual NAL systems}

Absorption systems are analyzed by means of the Monte Carlo Inversion 
(MCI) procedure 
(Levshakov et al. 2000, 2003; Agafonova et al. 2005, 2007) 
which is based on the assumption
that all lines observed in the metal system are formed in 
the same gas with fluctuating
density and velocity fields. The procedure's inputs are the observed 
line profiles and 
the ionization curves for each ion (hydrogen + metals) included in the analysis, 
the outputs~-- parameters of the gas density
and velocity distributions and individual metal abundances
which are supposed to be constant within the absorber.
The column
densities of hydrogen and all metal ions are calculated as well. 
The ionization curves are
computed with the photoionization code CLOUDY version 07.02.01 
(last described by Ferland et al. 1998) 
which in turn uses as an input the adopted (trial) SED.

For the present work the computational procedure was modified in a way 
to account for the incomplete coverage of the background light source. 
With $C$ standing for the covering factor (fraction of the background 
source of the continuum radiation occulted by the absorbing cloud) 
and basing on principles of the geometrical optics one
can easily determine 
the observed normalized intensity $I_\lambda$ 
within the absorption line profile as
\begin{equation}
 I_\lambda = C \exp(-\tau_\lambda) + (1-C)\; .
\label{E1}
\end{equation}
The apparent optical depth $\tau_\lambda$ is calculated
through Eq.~(4) in Levshakov et al. (2000). 
The covering factor is assumed to be the same for the whole cloud, 
i.e. it does not depend on 
$\lambda$ (or, equivalently, on the radial velocity $v$). 
However, different covering factors are allowed for different ions.

It is well known that NAL systems with incomplete 
coverage often show $\lambda$-depending 
covering factors which may be caused by the overlapping of clouds 
with different covering factors
or by the interaction of several physical processes. In these cases 
the estimation of accurate
column densities becomes problematic and in the present study such 
systems are not used for the SED recovery.

The SED of the radiation which ionizes the NAL system is 
reconstructed according to the iterative 
procedure described in Agafonova et al. (2005, 2007). 
The procedure includes
(1) the parameterization of the spectral shape by means of
a set of variables (called `factors'), (2) the choice of a
quantitative measure (called `response') to evaluate goodness
of a trial spectral shape, and (3) the estimation of a direction
in the factor space which leads to a spectrum with the better goodness.
Moving along this direction, we come from the initial spectral shape to
the one with better characteristics concerning the fitting 
of the observed line intensities.
Now the MCI calculations are carried out with this newly obtained
ionizing spectrum and the whole procedure is
repeated till the optimal spectral shape is found, i.e. one
which allows to reproduce the observed intensities of all lines
without any physical inconsistencies. 

All calculations throughout the paper were performed with laboratory 
wavelength and oscillator strengths taken from Morton (2003)
for $\lambda > 912$ \AA\ and from Verner et al. (1994) 
for $\lambda < 912$ \AA. Solar abundances were
taken from Asplund et al. (2004). 
Note that their solar abundances of neon, Ne/H = $7.5\times10^{-5}$,
and nitrogen, N/H = $6\times10^{-5}$, are ~1.4
times (0.15 dex) lower than those from Holweger (2001) which are used as default
inputs in CLOUDY.

\subsection{NAL systems towards \object{J 2233-606} }

The quasar \object{J 2233-606} (\zem = 2.238) 
was observed both at the VLT/UVES and HST/STIS with a wide
wavelength coverage from 2300 \AA\ to $\simeq$~10000 \AA\ 
(Cristiani \& D'Odorico 2000).  The spectrum reveals
an associated complex consisting of several absorption systems 
extending over 1000 \kms. 
Due to the wavelength coverage it becomes possible to identify not 
only the usually observed lines of the doublets 
\ion{C}{iv}$\lambda\lambda 1548,1551$, \ion{N}{v}$\lambda\lambda 1238,1242$, and
\ion{O}{vi}$\lambda\lambda 1032,1037$, but also the lines of 
\ion{Ne}{viii}$\lambda\lambda 770,780$  
(albeit in a very noisy part of the STIS spectrum). 

The associated systems towards \object{J 2233--606} have been studied
twice~-- by Petitjean \& Srianand (1999) and by Gabel et al. (2006). 
Both groups determined the column densities of different species, $N_i$,
using velocity-dependent covering factors and after that compared $N_i$ with the
values predicted by models where simple power law SEDs 
in the range from UV to X-ray regions were assumed. 
It was concluded that such models cannot account 
for the observed strong \ion{Ne}{viii} 
absorption and that an additional high-ionization absorber 
is needed to reproduce the \ion{Ne}{viii} line. 
In the present section
we re-analyze two absorption systems from this complex using 
physical model of the associated absorber as outlined above.
Namely, gas in the absorbing cloud has varying density and velocity
and the covering factor is assumed to be identical for all points in
the line profile (no velocity dependence). However, the covering
factor can differ for individual ions since different ions trace
different parts of the gas resulting in different solid angles
subtended.

\subsubsection{System at \zabs = 2.198}

Apart from metal doublets, this system 
exhibits also clear lines
of hydrogen Ly-$\alpha$ and Ly-$\beta$ (Fig.~\ref{fg_1}). 
There are continuum windows in the QSO spectrum at the expected positions
of \ion{C}{iii}$\lambda 977$ \AA\ and \ion{Si}{iv}$\lambda 1393$ \AA, the expected
position of \ion{O}{iv}$\lambda 787$ \AA\ is blended with Ly-$\gamma$ from the 
\zabs = 1.59 system  (\ion{O}{iv} profile is overplotted 
on the panel with
\ion{O}{vi}$\lambda 1031$ \AA\  in Fig.~\ref{fg_1}). 

First trial calculations were performed with the power law SED, $\alpha = 1.5$. 
The objective function included hydrogen
lines plus all lines of metal doublets mentioned above except 
for \ion{Ne}{viii} because of its too low S/N ratio. 
Since the system under consideration does
not show ionic lines of subsequent ionization stages, 
additional assumptions are required to fix the mean ionization parameter $U$. 
For oversolar metallicity which is reliably reproduced for 
this system independently on the model assumptions, the usual constraint  
is the solar relative abundance of oxygen to carbon, 
i.e. [C/O]~$\sim$0.

The observed profiles of all doublets can be well described by a model  
with a single covering factor which takes individual values for each ion 
(smooth curves in Fig.~\ref{fg_1}). 
The estimated covering factors along with the column densities of ions are 
given in Table~\ref{tbl-1}. 

It is to note that in the present case 
both the covering factors and the column densities are invariant and depend
neither on the assumed SED nor on the calculation method. 
For instance, a commonly used deconvolution of the line profiles into
subcomponents delivers the same values for the covering factors 
and column densities as obtained with the MCI procedure.

\begin{figure}[t]
\vspace{-0.5cm}
\hspace{-0.2cm}\psfig{figure=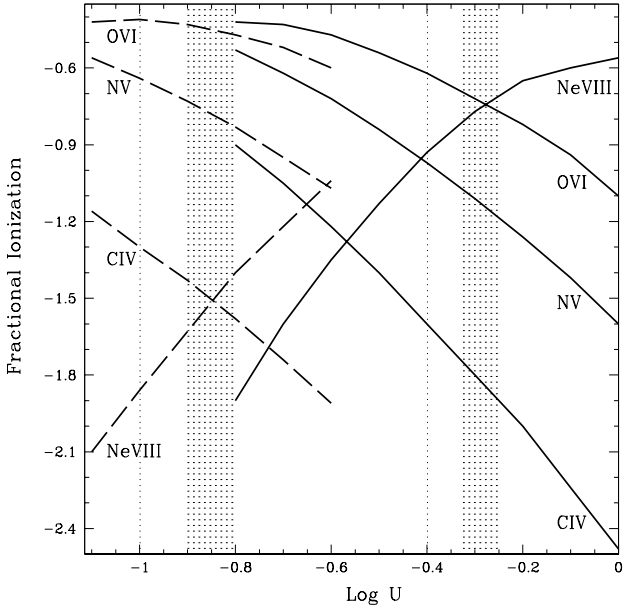,height=10cm,width=9.7cm}
\vspace{-1.2cm}
\caption[]{
Ionization fractions of different ions calculated for: 
{\it dashed lines}~-- power law
ionizing spectrum $F_\nu \propto \nu^{-1.5}$; 
{\it solid lines}~-- 
ionizing spectrum with a break at $\sim$4 Ryd 
(Fig.~\ref{fg_3}).
Calculations are performed with solar abundances 
and for the thermal and ionization equilibrium. See text for more
details.
}
\label{fg_2}
\end{figure}

For the adopted SED ($\nu^{-1.5}$)
the ionization curves for ions \ion{C}{iv}, \ion{O}{vi} and 
\ion{Ne}{viii} are plotted by dotted lines in Fig.~\ref{fg_2} 
with shadowed area indicating the 
range of the ionization parameter $U$ inside the absorber. 
Both neon and oxygen are
$\alpha$-elements and are expected to have the relative abundances 
[Ne/O]~$\sim$0. 
This gives a predicted column density
$N_{\scriptstyle\rm Ne\, {\scriptscriptstyle VIII}} \sim 5\times10^{13}$ \cm\ 
with the corresponding synthetic profile shown by
the dashed curve in the \ion{Ne}{viii} panel in Fig.~\ref{fg_1}. 
Obviously the line is too shallow to account for the 
observed \ion{Ne}{viii} absorption.
This is valid for all power law SEDs since the arrangement of \ion{C}{iv},
\ion{O}{vi} and \ion{Ne}{viii} ionization curves does not vary 
with increasing/decreasing spectral index: 
the curves are simply shifted along 
the $U$ axis to the right/left, but in the $U$ range
ensuring the ratio [C/O]~$\sim$0 the fraction 
of \ion{Ne}{viii} remains too small. 

\begin{figure}[t]
\vspace{0.0cm}
\hspace{-0.2cm}\psfig{figure=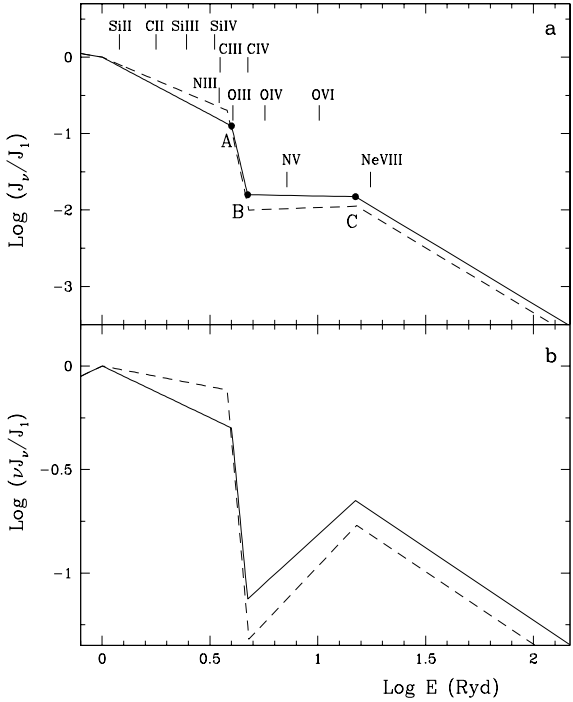,height=10cm,width=9.7cm}
\vspace{-1.0cm}
\caption[]{
Spectral energy distributions (panel {\bf a} intensity,
panel {\bf b} flux) of the quasar radiation
proposed for the associated systems towards \object{J 2233--606}
to explain the presence of \ion{C}{iv}, \ion{O}{vi}, and \ion{Ne}{viii}.
The spectra are normalized so that $J_\nu(h\nu =$ 1 Ryd) = 1.
The solid line between 1 Ryd and point $A$ corresponds to $\alpha = 1.5$,
the dashed line within the same region~-- to $\alpha = 1.2$.  
}
\label{fg_3}
\end{figure}

The situation changes for a SED with a break around 4 Ryd like the one
plotted by solid line in Fig.~\ref{fg_3}. 
The ionization curves for this SED
are given in the right hand part of Fig.~\ref{fg_2} (solid lines) 
with the shadowed area indicating the range of the fractions
of \ion{C}{iv} and \ion{O}{vi} ensuring [C/O]~$\sim$0. 
Now the fraction of \ion{Ne}{viii} is nearly
equal to that of \ion{O}{vi} leading to the column density 
of \ion{Ne}{viii} an order
of magnitude higher than from the power law SED: 
$N_{\scriptstyle\rm Ne\, {\scriptscriptstyle VIII}} \sim 5\times10^{14}$ \cm.
The corresponding synthetic profile for \ion{Ne}{viii} is plotted 
in Fig.~\ref{fg_1} for the covering factor 
$C_{\scriptstyle\rm Ne\, {\scriptscriptstyle VIII}} = 0.9$ (smooth line). 
A noisy HST/STIS spectrum at the positions of
\ion{Ne}{viii}$\lambda\lambda 770,780$ \AA\  
prevents accurate determination of both the covering factor and
column density, but it is clearly seen that the adopted~--
broken power law~--  SED produces 
$N_{\scriptstyle\rm Ne\, {\scriptscriptstyle VIII}}$  
consistent with observations. 
This SED gives also the following abundances: 
[C,O,Ne/H]~= 0.5, [N/H]~= 0.9-1.0, i.e.
compared to the solar values nitrogen
is 2.5-3 times overabundant than other elements.

\begin{table}[t!]
\centering
\caption{Parameters of the NAL system \zabs = 2.198 
towards \object{J 2233--606} }
\label{tbl-1}
\begin{tabular}{ccc}
\hline
\hline
\noalign{\smallskip}
Ion & $N_i$, cm$^{-2}$ & $C_i$ \\
\noalign{\smallskip}
\hline 
\noalign{\medskip}
H\,{\sc i} & $(1.1\pm0.1)$E14 & $0.60\pm0.03$\\
C\,{\sc iv} & $(1.7\pm0.1)$E14 & $0.72\pm0.01$\\
N\,{\sc v} & $(5.3\pm0.8)$E14 & $0.70\pm0.02$\\
O\,{\sc vi} & $(3.4\pm0.7)$E15 & $0.82\pm0.02$\\
\noalign{\smallskip}
\hline
\end{tabular}
\end{table}

The absence of low-ionization lines and unknown metallicity of the
\zabs=2.198 absorber does not allow us to
recover all details of the underlying SED of the ionizing radiation. 
Numerical simulations with different
types of SED show that the
simultaneous existence of significant amounts 
of \ion{C}{iv}, \ion{O}{vi} and \ion{Ne}{viii} can be
provided only by a sharp break in the spectrum around the energy of 4 Ryd. 
However, other spectral shape features are determined with much less accuracy.
For instance, positions of the ionization curves shown in Fig.~\ref{fg_2} 
are insensitive to the spectral index $\alpha$ in the energy range below the
discontinuity point $A$ (shown in Fig.~\ref{fg_3}) at
1 Ryd\ $< E < 4$ Ryd.

A deeper break after point $A$ does not affect the ionization curve 
of hydrogen, but shifts
the ionization curves of ions parallel to the right. 
The condition [C/O]~$\sim$0 is now fulfilled at higher $U$ with
correspondingly lower fraction of neutral hydrogen~-- thus, 
we obtain the same column density of \ion{Ne}{viii}
but the metal content becomes lower: the dashed line SED in Fig.~\ref{fg_3} gives
metallicity [C,O,Ne/H]~= 0.18 with stable overabundance of nitrogen by 0.5 dex. 
Power law index between points $B$ and $C$ can vary from $\alpha = -0.3$ to 
$\alpha = 0.1$ 
without significant changes in the predicted 
$N_{\scriptstyle\rm Ne\, {\scriptscriptstyle VIII}}$, 
but harder spectra deliver
more \ion{Ne}{viii} and are probably preferable. 
The discontinuity point $C$ and the power law index beyond it
are set in a way to account for measurements of AGN/QSO luminosity 
at $E$~= 2 keV~= 147 Ryd (Steffen et al. 2006):
again, metal lines available in the \zabs = 2.198 system are weakly sensitive 
to the coordinate of this point.

\begin{figure}[t]
\vspace{-0.7cm}
\hspace{-0.2cm}\psfig{figure=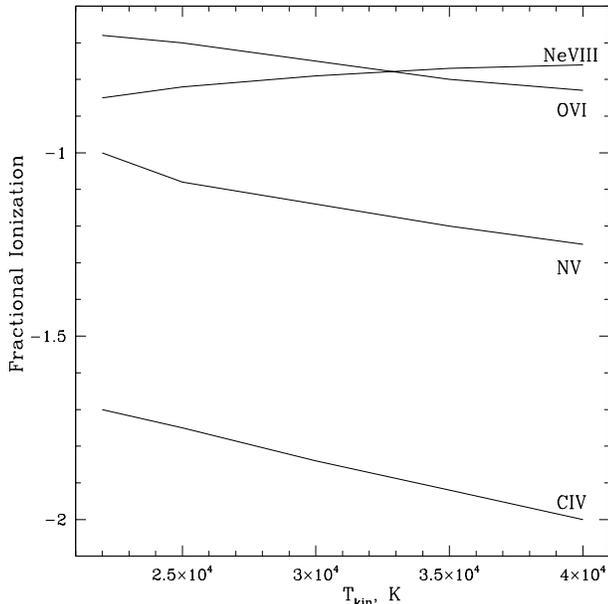,height=10cm,width=9.7cm}
\vspace{-1.5cm}
\caption[]{
Ionization fractions (in $\log_{10}$) of different ions versus kinetic temperature
calculated for the incident ionizing spectrum shown in
Fig.~\ref{fg_3} (solid line)
at the ionization parameter $U = 0.45$. For the solar metallicity the 
equilibrium temperature is 22000~K. 
}
\label{fg_4}
\end{figure}

All calculations above were performed assuming gas in thermal equilibrium. 
However, conditions in the
vicinity of the central engine vary rapidly, and gas can become overheated. 
Fig.~\ref{fg_4} shows the dependence of the ionization
fractions of different ions on the temperature calculated for the UV
spectrum from Fig.~\ref{fg_3} (solid line)
and the ionization parameter $U = 0.45$ which for the solar metallicity 
gives the equilibrium gas temperature of 22000~K. 
While the fractions of \ion{C}{iv} and \ion{O}{vi} decrease with rising $T$, 
the fraction of \ion{Ne}{viii} increases, 
i.e. the overheated gas is more abundant in \ion{Ne}{viii}. 
On the other hand, the cooling time is determined
by the gas density, and, hence, the rarefied volumes~-- those traced 
by \ion{Ne}{viii}~-- remain longer in the
overheated state. Thus, a fraction of \ion{Ne}{viii} at the value of $U$ 
corresponding to [C/O]~$\sim$0 will be higher 
in the cooling gas than that at the equilibrium.

To summarize, the
ionizing spectrum with a sharp break in the intensity around 
4 Ryd can easily provide conditions for
significant amounts of \ion{C}{iv}, \ion{O}{vi} and \ion{Ne}{viii} 
to arise in the same absorbing cloud.

The results obtained allow us also to make some conclusions concerning 
the physical state of the absorbing cloud itself. Firstly, the ratio 
$N_{\scriptstyle\rm C\, {\scriptscriptstyle IV}}/
N_{\scriptstyle\rm O\, {\scriptscriptstyle VI}} = 20$ along with the 
constraint [C/O]~$\sim$0 and the absence (or small amount) 
of \ion{O}{iv} (Fig.~\ref{fg_1}) unambiguously
point to photoionization as the source of the observed ionization state: 
in case of collisional ionization the leading oxygen ion at the
temperature which corresponds to 
$N_{\scriptstyle\rm C\, {\scriptscriptstyle IV}}/
N_{\scriptstyle\rm O\, {\scriptscriptstyle VI}} = 20$
is \ion{O}{iv} with column density an order of magnitude 
larger than 
$N_{\scriptstyle\rm O\, {\scriptscriptstyle VI}}$ 
(cf., Sutherland \& Dopita 1993). 
Previously it was noted by

The second conclusion concerns the ion-dependent covering factors. 
For the system under study this dependence
was stated previously by Petitjean \& Srianand (1999), and they explained 
this fact by different distances to the broad emission line region. 
However, distribution of covering factors as shown in Table.~\ref{fg_1} directly
follows from a model of the associated absorption arising in a gas cloud with 
fluctuating density. Namely, ions of
higher ionization stages trace rarefied gas which can be quite extended 
whereas low ionization ions
originate in more dense and, hence, compact volumes. 
This naturally explains the increasing value of the
covering factor from \ion{H}{i} to \ion{O}{vi} and the reason
why the synthetic \ion{Ne}{viii} profile
in Fig.~\ref{fg_1} was plotted with 
$C_{\scriptstyle\rm Ne\, {\scriptscriptstyle VIII}} = 0.9$.

\subsubsection{System at \zabs = 2.205}

\begin{figure*}[t]
\vspace{0.0cm}
\hspace{-0.2cm}\psfig{figure=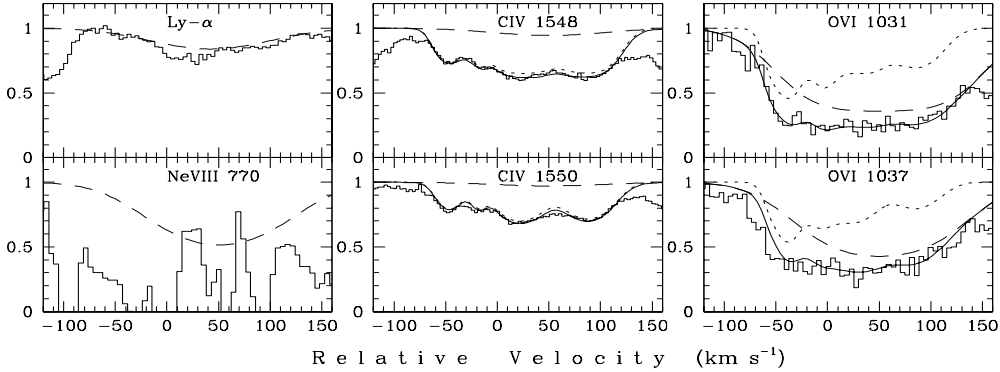,height=16cm,width=18cm}
\vspace{-10.5cm}
\caption[]{Same as Fig.~\ref{fg_1} but
for the \zabs = 2.205 system
towards \object{J 2233--606} (solid-line histograms).
The zero radial velocity is fixed at $z = 2.2050$.
Synthetic profiles of \ion{C}{iv} and \ion{O}{vi} 
are produced by broad and shallow absorption
(dashed curves) with covering factor $C \sim$0.6, and by several 
overlapped narrow absorptions (dotted curves) with $C \sim$0.3.
Smooth curves represent the convolution of all components according to
Eq.(\ref{E2}). 
The dashed line in the \ion{Ne}{viii} panel~-- predicted profile for the broad
component corresponding to the
ionizing spectrum with a break at $\sim$4 Ryd (Fig.~\ref{fg_3}, solid line).
The dashed line in the Ly-$\alpha$ panel~-- predicted upper limit for the
broad component.
}
\label{fg_5}
\end{figure*}

This portion of the associated absorption complex differs  
from the system considered above: profiles of metal doublets are 
more shallow and broad, column densities of \ion{C}{iv}
and \ion{O}{vi} seem to be not very different, but the amount
of \ion{Ne}{viii} is nevertheless significant (Fig.~\ref{fg_5}).
Preliminary calculations have shown that profiles of the
\ion{C}{iv}$\lambda\lambda 1548,1550$ \AA\ lines
in this system cannot be fitted with a single covering factor. 
Winding shape of the profiles
indicates that the observed absorption may be caused by several gas clouds, 
each with its own covering factor.
For a simplest model with multiple cloud overlapping the observed 
intensity can be written as
\begin{equation}
 I_v = \prod\limits^n_{k=1} \left( 1 - C_{i,k} + C_{i,k} {\rm e}^{-\tau_k(v)}
\right)\ ,
\label{E2}
\end{equation}
where $C_{i,k}$ is a covering factor of cloud $k$ for ion $i$,
and $v$ is radial velocity.

The unsaturated and unblended \ion{C}{iv} lines can be deconvolved 
into separate subcomponents. Calculations
give one broad component with the Doppler parameter $b = 74$ \kms\ and
covering factor $C_{\rm broad} \sim$0.6 
and four narrow components with $b = $10-26 \kms\ and covering factors 
$C_{\rm narrow} = $0.2-0.35.
The column density of the broad component is  
(2-4)$\times10^{13}$ \cm, which is only a small fraction 
of a joint column density $5.5\times10^{14}$ \cm\ of the
narrow components (Table~\ref{tbl-2}). 
Synthetic profiles of the broad and narrow components as well as the
combined line profile are plotted in Fig.~\ref{fg_5}. 
Since the profiles of
\ion{O}{vi} are saturated and smooth they
can be deconvolved only under additional constraints. 
Assuming the same velocity and component structure
as for \ion{C}{iv} doublet, we obtain the column density for 
the broad \ion{O}{vi} component $\sim$$1.5\times10^{15}$ \cm\ 
and similar column density for the combined
narrow components (Table~\ref{tbl-2}). Thus,
the broad component shows the ratio 
$N_{\scriptstyle\rm O\, {\scriptscriptstyle VI}}/
N_{\scriptstyle\rm C\, {\scriptscriptstyle IV}} \sim$40-70 
which points to a highly ionized gas with the
ionization parameter $U \sim$1 (Fig.~\ref{fg_2}), 
whereas the narrow components have 
$N_{\scriptstyle\rm O\, {\scriptscriptstyle VI}}/
N_{\scriptstyle\rm C\, {\scriptscriptstyle IV}} \sim$3 
and are much lower ionized with $U \sim$0.1. 
With these parameters the predicted column density for 
\ion{Ne}{viii} is $\sim$$6\times10^{14}$ \cm\
for the broad component 
(plotted in Fig.~\ref{fg_5} for the covering factor 0.7) 
and $\sim$$10^{13}$ \cm\ for the narrow components.

The line \ion{N}{v}$\lambda 1238$ \AA\ is blended with 
\ion{Ca}{ii}$\lambda 3968$ \AA\ from \zabs~$\sim$0 which hampers
its deconvolution into components and estimation of their covering factors. 
However, trial calculations
with different constraints show that 
an overabundance of nitrogen is present 
in both broad and narrow components.

\begin{table}[t!]
\centering
\caption{Column densities and covering factors for the broad (b) and
four narrow (n) components of the associated system \zabs = 2.205 
towards \object{J 2233--606} shown in Fig.~\ref{fg_5} }
\label{tbl-2}
\begin{tabular}{ccccc}
\hline
\hline
\noalign{\smallskip}
Ion & $N_i$, cm$^{-2}$ & $\sum N_i$, cm$^{-2}$ & $C_i$ & $C_i$ \\[-2pt]
    & {\footnotesize (b)} & {\footnotesize (n)} &
      {\footnotesize (b)} & {\footnotesize (n)} \\ 
\noalign{\smallskip}
\hline 
\noalign{\medskip}
H\,{\sc i} & $\la 3$E13 & & $\la 0.5$ \\
C\,{\sc iv} & $(2-4)$E13 & $(5.5-5.7)$E14 & 0.6 & 0.2--0.35 \\
O\,{\sc vi} & $(1.5-1.7)$E15 & $(1.7-1.9)$E15 & 0.65 & 0.3--0.4\\
Ne\,{\sc viii} & $(6-7)$E14 & & 0.7\\
\noalign{\smallskip}
\hline
\end{tabular}
\end{table}

\subsection{NAL systems towards \object{HE 1341--1020} }

The spectrum of the quasar \object{HE 1341--1020} 
was obtained with the UVES/VLT in the framework of the ESO Large
Program `QSO Absorption Line Systems' (ID No.166.A-0106). 
Data reduction was performed by B.~Aracil.

\object{HE 1341--1020} is a mini-BAL quasar with broad absorptions 
in \ion{C}{iv}, \ion{N}{vi} and \ion{O}{vi} extending over 3000 \kms. 
There is also a narrow absorption system at \zabs = 2.147. 
The quasar emission redshift estimated from the 
\ion{C}{ii}$\lambda 1334$ \AA\ 
and \ion{C}{iii}$\lambda 1990]$ \AA\ emission
lines is \zem~= $2.1485 \pm 0.002$.

\subsubsection{System at \zabs = 2.147}

\paragraph{\it Ionization state and restored SED.}

\begin{figure*}[t]
\vspace{0.0cm}
\hspace{-0.2cm}\psfig{figure=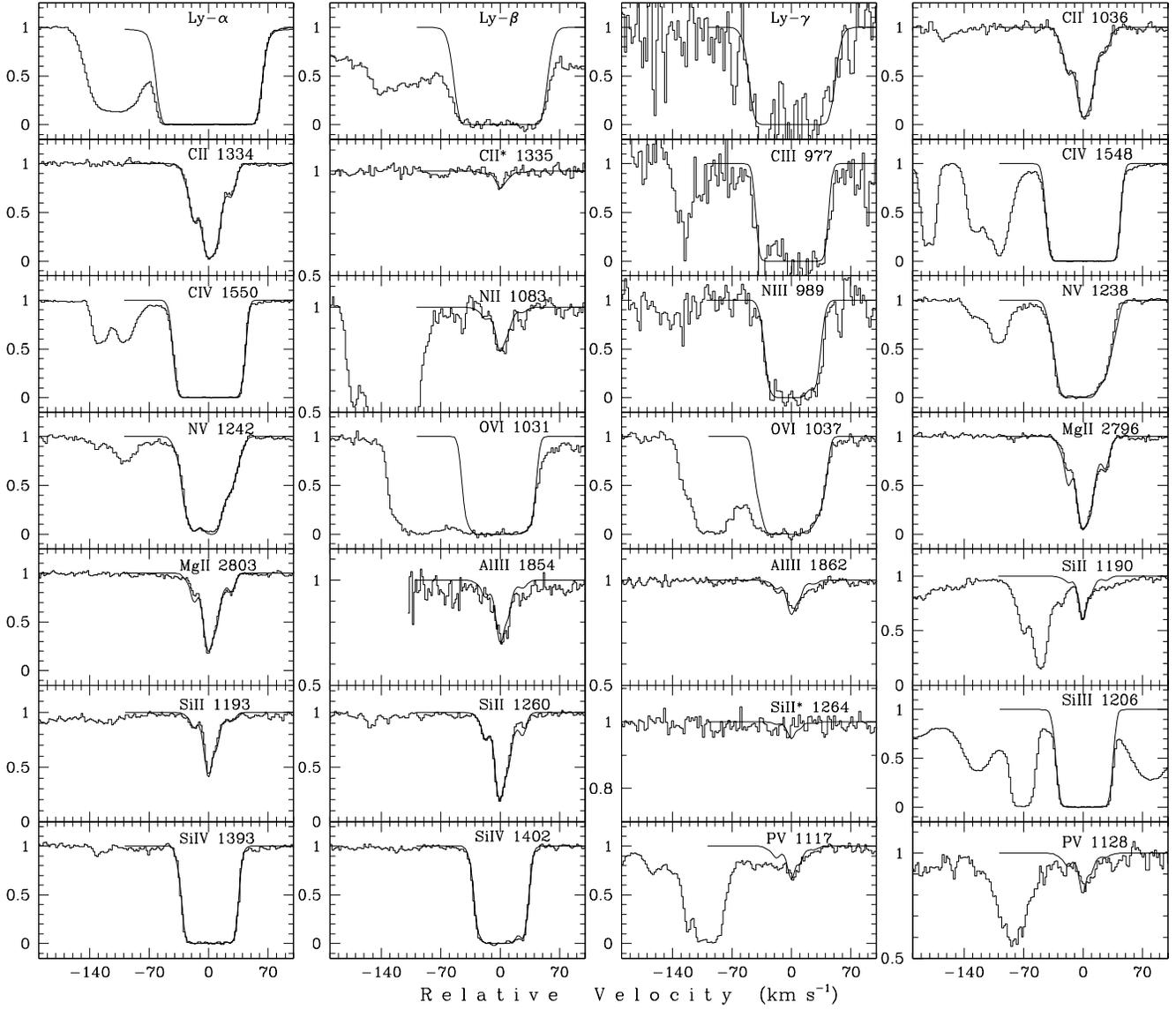,height=18cm,width=18cm}
\vspace{-2.5cm}
\caption[]{Same as Fig.~\ref{fg_1} but for
the \zabs = 2.147 system
towards \object{HE 1341--1020} (solid-line histograms).
The zero radial velocity is fixed at $z = 2.14736$.
Synthetic profiles corresponding to the
ionizing spectrum with a break at $\sim$4 Ryd (Fig.~\ref{fg_7})
are plotted by smooth curves.
Note different vertical scales in some panels.
}
\label{fg_6}
\end{figure*}

The system at \zabs = 2.147 exhibits a wealth of lines of metal 
ions ranging from low ionization \ion{Si}, \ion{N}{ii},
\ion{Mg}{ii} up to high ionization \ion{N}{v} and \ion{O}{vi} (Fig.~\ref{fg_6}).
The presence of many elements in different stages of ionization makes 
it possible to reconstruct the shape
of the \object{HE 1341--1020} spectrum in more detail than that previously
described for \object{J 2233--606}.
Trial calculations 
revealed that power law spectra as well as
the AGN spectrum of Mathews \& Ferland (1987) failed 
to reproduce the observed ratio \ion{N}{ii}/\ion{N}{v}
resulting in the synthetic line profiles which significantly underestimate
the observed intensity of  
\ion{N}{ii}$\lambda 1083$ \AA, 
and yielded relative element abundances which did not
comply with existing observational and theoretical data
(e.g., [Mg/O]~$ \simeq 0.7$, [Si/O]~$ \simeq 0.3$). 
Several types of SEDs with the break at 4 Ryd were tried 
as well and as a result the SED recovered from the \zabs = 1.78
associated system towards \object{HE 0141--3432} (Reimers et al. 2005,
Sect.~4.2) was chosen as the best initial guess for the spectral
shape recovering procedure.
The search for an optimal spectral
shape was directed by the requirement to reproduce all observed 
lines under the constraint 
[\ion{Si},\ion{Mg}, \ion{O}/\ion{C}]~$ \la 0.2$. 
Details of calculations are given in Reimers et al. (2005, appendix).

\begin{figure}[t]
\vspace{0.0cm}
\hspace{-0.2cm}\psfig{figure=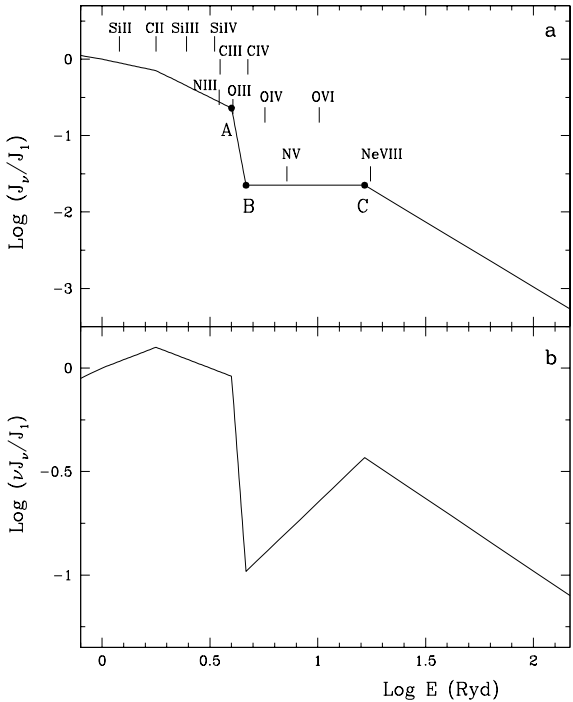,height=10cm,width=9.7cm}
\vspace{-1.0cm}
\caption[]{SED estimated from the \zabs = 2.147 system towards 
the quasar \object{HE 1341--1020} (\zem = 2.1485).
}
\label{fg_7}
\end{figure}

The resulting SED is shown 
in Fig.~\ref{fg_7}, and the corresponding synthetic profiles
are plotted by solid curves in Fig.~\ref{fg_6}.  The estimated
physical parameters and column densities are given in Table~\ref{tbl-3}.

As it was for the previous system, the spectral shape again has 
a sharp break at $\sim$4 Ryd, but
in the present case it is also possible to fix the depth of this break 
and the broken power law slope between 1 and 4 Ryd. 

The column density of neutral hydrogen is 
$N_{\scriptstyle\rm H\,{\scriptscriptstyle I}} \la 10^{17}$ \cm, i.e. 
the \zabs = 2.147 system is a sub-LLS (optical depth in the 
\ion{H}{i} Lyman continuum $\tau^{\rm H\,{\scriptscriptstyle I}}_c \la 0.6$). 
The predicted column density for \ion{He}{ii} is 
$N_{\scriptstyle\rm He\,{\scriptscriptstyle II}} \la 3\times10^{18}$ \cm\
which makes the system moderately optically thick to the 
\ion{He}{ii} Lyman continuum 
($\tau^{\rm He\,{\scriptscriptstyle II}}_c \simeq 4.7$).

The ionizing spectrum is reconstructed from the observed absorption lines and 
therefore is sensitive to the local
processes in the absorbing cloud itself 
(for more details, see Sect.~3 in Agafonova et al. 2005). 
In particular, for the case in question the restored quasar spectrum 
can be affected by the \ion{He}{ii} Lyman continuum absorption
inside the cloud, i.e. the incident SED can be harder (probably by 0.2-0.3 dex) 
at energies $E > 4$ Ryd than the spectrum shown in Fig.~\ref{fg_7}.  
The \ion{He}{ii} Lyman continuum absorption is accompanied
by the recombination emission line of \ion{He}{ii} Ly-$\alpha$ (304 \AA)
and by the two-photon emission which both produce a characteristic emission
feature at $E \la 3$ Ryd in the transmitted spectrum of the incident
quasar radiation.  
Whether such a feature is present or not in 
the restored spectrum cannot be stated unambiguously 
because of the saturated profiles of the \ion{Si}{iii} and
\ion{Si}{iv} lines (fractions of silicon ions are most 
sensitive to the spectral shape of the ionizing radiation 
at $E \la 3$ Ryd as seen in Fig.~2
in Agafonova et al. 2007).  

The measured
metallicity of the absorbing gas is solar (carbon) or slightly 
oversolar ($\alpha$-elements) with
clear underabundances of nitrogen (0.6 solar) and aluminium (0.3 solar).

\paragraph{\it Distance to the QSO.}
The presence of the ground fine-structure line 
\ion{C}{ii}$^{\ast} \lambda 1335$ \AA\ arising from the first excited 
$^3{\rm P}^e_1$ level provides a rare opportunity
to measure from the photoionization model the mean gas number density and, 
hence, the distance to the light source.
For an ion in the interstellar (intergalactic) medium, the ratio
of excited ($n_2$) to ground-state ($n_1$)
population is equal to the ratio of the
collisional excitation rate $Q_{1\rightarrow2}$ to the spontaneous
transition probability $A_{2\rightarrow1}$ (Bahcall \& Wolf 1968):
\begin{equation}
\frac{n_2}{n_1} = \frac{Q_{1\rightarrow2}}{A_{2\rightarrow1}}\: .
\label{E3}
\end{equation}
The atomic data for \ion{C}{ii}$^\ast$ are the following
(Silva \& Viegas 2002): 
$A_{2\rightarrow1} = 2.291\times10^{-6}$ s$^{-1}$, 
the excitation rate by collisions with electrons at $T_{\rm kin} = 10^4$~K
$q^{\rm e}_{1\rightarrow2}
\simeq 1\times10^{-7}$ cm$^3$ s$^{-1}$. 
Since collisions with other particles have much lower excitation rates,
we put $Q_{1\rightarrow2} = q^{\rm e}_{1\rightarrow2}\,n_{\rm e}$.

From the measured column densities 
$N_{\scriptstyle\rm C\,{\scriptscriptstyle II}}$ and
$N_{\scriptstyle\rm C\,{\scriptscriptstyle II}^\ast}$ (see Table~\ref{tbl-3})
we obtain the ratio of the mean values $n_2$ to $n_1$
$$
\frac{n_2}{n_1} \simeq 0.02\, .
$$ 

Since this system has a high degree of ionization
($N_{\scriptstyle\rm H\,{\scriptscriptstyle I}}/
N_{\scriptstyle\rm H} = 
n_{{\scriptstyle\rm H}^+}/n_{\scriptstyle\rm H} \gg 1$, see Table~\ref{tbl-3}),
and collisions with other particles have much lower excitation rates, 
the total gas density equals to $n_0 \simeq 0.5$ \cmm\
(the contribution of
the ionized helium is ignored since it has a small effect).

If this system would be an intergalactic absorber, then
the ionization parameter $U = 0.02$ would correspond to the gas number density
$n_0 \simeq 0.001$ \cmm (assuming the intensity $J_{912}$ of the metagalactic 
UV background radiation at
$z = 2.0$ as given by Haardt \& Madau, 1996).
This is a clear indication that the ionizing 
background at the position of the \zabs = 2.147 system is enhanced 
by more than 2 orders of magnitude as
compared to the intergalactic background. 

In order to estimate the distance to the light source
the QSO continuum luminosity
at the hydrogen Lyman limit ${\cal L}_{\nu_c}$ is to be known.
The intrinsic luminosity ${\cal L}_{\nu_c}$ can be determined
from the comparison of $V = 17.1$  
with the
specific flux of a star having $m_{\rm V} = 0.0$ outside the Earth's atmosphere,
$F_{\scriptstyle \rm V}^\ast =
3.8\times10^{-20}$ ergs cm$^{-2}$ s$^{-1}$ Hz$^{-1}$~:
$$
f_{\scriptstyle \rm V} =
10^{-0.4\cdot17.1}\, F_{\scriptstyle \rm V}^\ast = 5\times10^{-27}\,
{\rm ergs}\ {\rm cm}^{-2}\ {\rm s}^{-1}\ {\rm Hz}^{-1}. 
$$
With the approximation for luminosity distance $d_{\scriptstyle\rm L}$ 
proposed in Riess et al. (2004), 
one obtains $d_{\scriptstyle\rm L} \simeq 13$~Gpc and
the apparent monochromatic luminosity 
$$
{\cal L}_\nu = 4\pi\ d_{\scriptstyle\rm L}^2\ 
f_{\scriptstyle \rm V}\ (1+z)^{-1} \simeq 3\times10^{31}\,
{\rm ergs}\ \ {\rm s}^{-1}\ {\rm Hz}^{-1}\, ,
$$
which results in the luminosity near the Lyman limit
$$
{\cal L}_{\nu_c} =  {\cal L}_\nu \left( \frac{1750}{912} \right)^{-1}
\simeq 1.5\times10^{31}\
{\rm ergs}\ \ {\rm s}^{-1}\ {\rm Hz}^{-1}\, .
$$ 
Given the gas number density $n_{\rm 0}$, 
the distance $r$ from the QSO to the absorbing cloud
can be calculated from the ionization parameter $U_0$
which is defined as
\begin{equation}
U_0 = \frac{Q({\rm H}^0)}{4\pi\, r^2\,c\,n_0} =
\frac{n_{\rm ph}}{n_0} \; , 
\label{E6}
\end{equation}
where 
\begin{equation}
Q({\rm H}^0) = \int^\infty_{\nu_c}\,\frac{{\cal L}_\nu}{h\,\nu}\,d\nu
\label{E7}
\end{equation}
is the number of hydrogen ionizing photons emitted per 
unit time by the central source. 

With the estimated above Lyman continuum luminosity
${\cal L}_{\nu_c}$, one finds
$Q({\rm H}^0) \simeq 2\times10^{57}$ photons s$^{-1}$,
assuming
${\cal L}_\nu = {\cal L}_{\nu_c}\,(\nu/\nu_c)^{-1}$. 
A substitution of the numerical values in (\ref{E6}) provides
$r \simeq 240$ kpc.
Accounting for the optical depth $\tau \la 0.6$ at the hydrogen
ionization edge due to absorption within the cloud, the distance
can be reduced to $\sim$170 kpc.

It is unclear whether an absorber at such a large distance from 
the light source can be gas ejected from the
QSO host galaxy.
However, distances as large as hundreds of kiloparsecs 
(mostly estimated as lower limits) are not exceptional
for the associated systems 
(Morris et al. 1986; Tripp et al. 1996; D'Odorico et al. 2004;
Reimers et al. 2005).

\subsubsection{System at \zabs = 2.107}

\begin{figure*}[t]
\vspace{0.0cm}
\hspace{-0.2cm}\psfig{figure=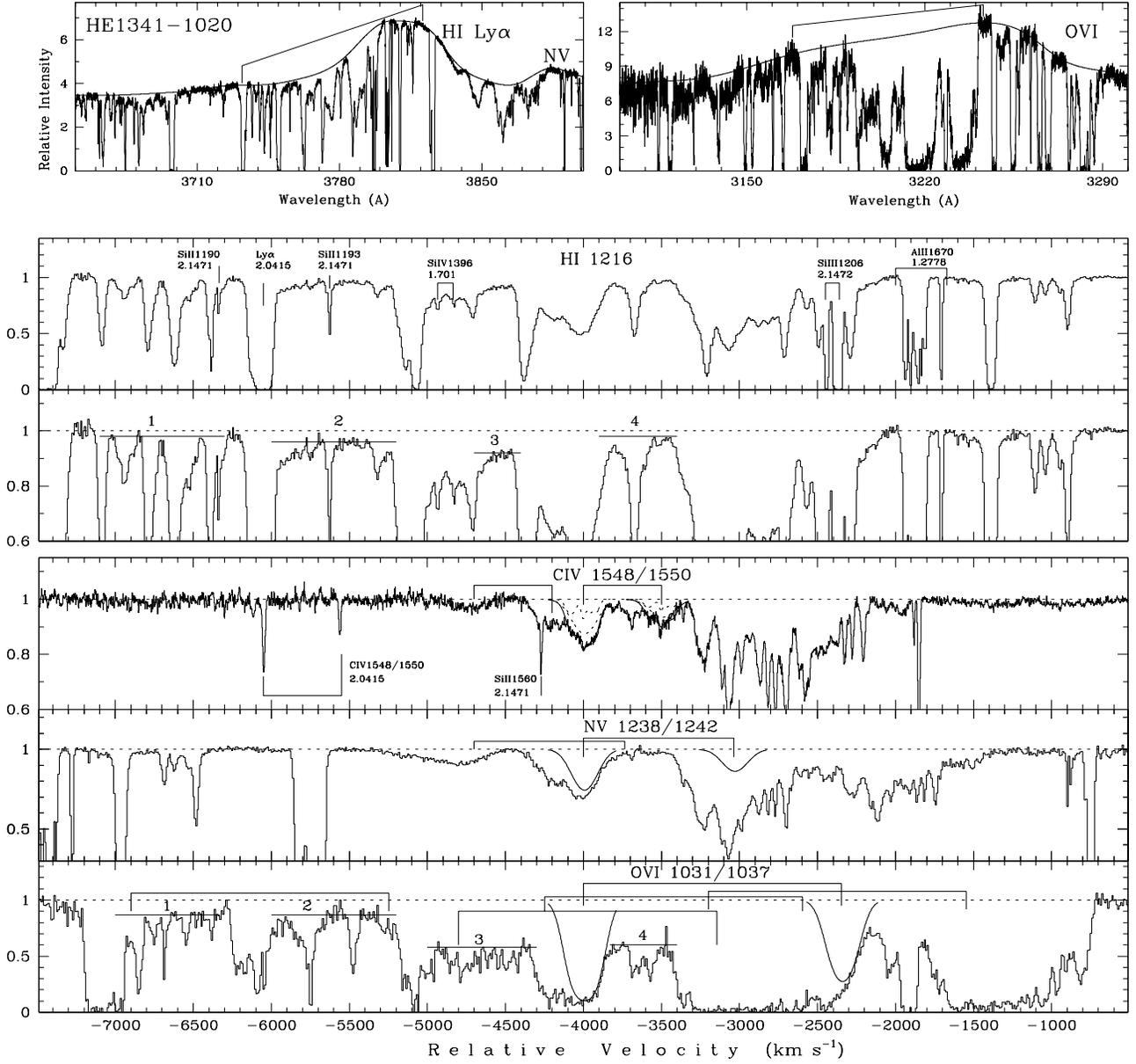,height=16.7cm,width=18cm}
\vspace{-0.25cm}
\caption[]{Mini-BAL absorptions towards \object{HE 1341--1020}.
Two upper panels represent portions of the original QSO spectrum
in the vicinity of broad emission lines \ion{H}{i} Ly-$\alpha$ and
\ion{O}{vi}+\ion{H}{i} Ly-$\beta$. The smooth curves here are the
local continua calculated through the spline interpolation between
'clear' continuum windows seen at high spectral resolution 
($FWHM \simeq$6 \kms) in the UVES/VLT data. The lower panels
show the zoomed portions marked by long brackets in the upper panels.  
The vertical axis is normalized intensities.
The zero radial velocity is fixed at \zem = 2.1485.
Two combined panels with \ion{H}{i} $\lambda 1216$ profiles are,
respectively, the general view with labeled absorption lines
identified at different redshifts and the same but zoomed in the vertical
scale spectrum to illustrate broad and shallow absorption features
(marked by numbers over horizontal lines) caused by \ion{H}{i} Ly-$\alpha$ 
arising from a fast gas outflow. The corresponding broad and shallow 
features are seen in the \ion{O}{vi} panel.
Absorption centered at $v = -4000$ \kms (\zabs = 2.107)
can be deconvolved into one broad and shallow component and
several overlapped narrow components (shown by dotted curves).
Smooth curves show: \ion{C}{iv}~-- synthetic profile of the convolved
broad and narrow absorptions; \ion{N}{v}~-- broad component corresponding
to the covering factors and the velocity structure as determined from
\ion{C}{iv}; \ion{O}{vi}~-- limiting profile for the broad component.  
Long brackets in the three bottom panels mark components of the corresponding
doublets to illustrate their strong overlapping.
}
\label{fg_8}
\end{figure*}

Mini-BAL absorptions in \ion{C}{iv}, \ion{N}{v} and \ion{O}{vi} 
at $-7000$ \kms\ $< v < -1000$ \kms\ 
are formed by a large number of
overlapping components (Fig.~\ref{fg_8}) which in most cases do not allow 
to estimate covering factors and, hence, to measure accurately   
column densities. Fortunately, portions of 
the \ion{C}{iv}$\lambda 1548$ \AA\ 
profile between $-4100$ \kms and $-3855$ \kms (\zabs~$\simeq 2.1066$) and, correspondingly,  
\ion{C}{iv}$\lambda 1550$ \AA\ profile at 
$-3638$ \kms $< v < -3396$ \kms are not blended
and can be used for quantitative analysis. 
A winding  shape of the line profiles favors the  
model used above for the \zabs = 2.205 system towards \object{J 2233-606}: 
several narrow components seen against 
the broad one. Component fitting according to Eq.(\ref{E2}) gives for 
the broad component the column density
$N_{\scriptstyle \rm C\,{\scriptscriptstyle IV}, broad} = 
4.2\times10^{13}$ \cm\ and the covering factor
$C_{\scriptstyle \rm C\,{\scriptscriptstyle IV}, broad} = 0.8$, 
and for the four narrow components the total
column density
$N_{\scriptstyle \rm C\,{\scriptscriptstyle IV}, narrow} = 
5.8\times 10^{13}$ \cm\ and the covering factor  
$0.16 \la C_{\scriptstyle \rm C\,{\scriptscriptstyle IV}, narrow} \la 0.36$ 
(thin smooth and dotted curves in panel \ion{C}{iv} in Fig.~\ref{fg_8}).
 
From the doublets \ion{N}{v} and \ion{O}{vi} 
only the lines \ion{N}{v}$\lambda 1238$ \AA\ and
\ion{O}{vi}$\lambda 1031$ \AA\ are available. 
Assuming for \ion{N}{v} 
similar velocity structure and
covering factors as calculated for \ion{C}{iv}, one obtains 
for the broad \ion{N}{v} component the
column density 
$N_{\scriptstyle \rm N\,{\scriptscriptstyle V}, broad} = 
(1.4-1.6)\times10^{14}$ \cm. 
The saturated  \ion{O}{vi}$\lambda 1031$ \AA\
line indicates the covering factor of 
$C_{\scriptstyle \rm O\,{\scriptscriptstyle VI}, broad} \simeq 0.90-0.95$, 
which gives the upper limit for the column density
of the broad \ion{O}{vi} component 
$N_{\scriptstyle \rm O\,{\scriptscriptstyle VI}, broad}  < 2\times10^{15}$ \cm.

The ratio 
$N_{\scriptstyle \rm O\,{\scriptscriptstyle VI}, broad}/
N_{\scriptstyle \rm C\,{\scriptscriptstyle IV}, broad} \la 50$
together with the assumption of solar relative abundances [C/O]~$\sim$0 
lead to the range of the ionization parameters $U$ 
which translate the column densities of 
\ion{C}{iv} and \ion{N}{v} observed in the broad
component into the reliable relative overabundance of nitrogen [N/C]~$ > 0$ 
for both a power law SED and a SED with a discontinuity at 4 Ryd.
Thus, the gas comprising the
broad component in the mini-BAL absorption at 
the radial velocity $v  \simeq -4000$ \kms obviously 
differs from  the gas producing the narrow line
absorption at \zabs = 2.14736 (Fig.~\ref{fg_6}) where the nitrogen 
underabundance of [N/C]~$= -0.15$ was detected.
This will be discussed further in Sect.~3.2.

\subsection{NAL systems towards \object{HE 2347--4243} }

An associated complex towards the quasar 
\object{HE 2347--4243} (\zem = 2.902) consisting 
of many absorption systems and
extending over 1500 \kms was described in detail in Fechner et al. (2004). 
It was found that  column
densities of ions measured in certain systems favored the ionizing spectrum 
with a break at $E \sim$4 Ryd.
However, an exact spectral shape in the range $E > 1$ Ryd was not estimated. 
Below we re-analyze a few systems from
this complex with the objective to reach more tenable conclusions 
about the SED of \object{HE 2347--4243}.

\subsubsection{System at \zabs = 2.898}
\label{Ss-5}

\begin{figure*}[t]
\vspace{0.0cm}
\hspace{-0.2cm}\psfig{figure=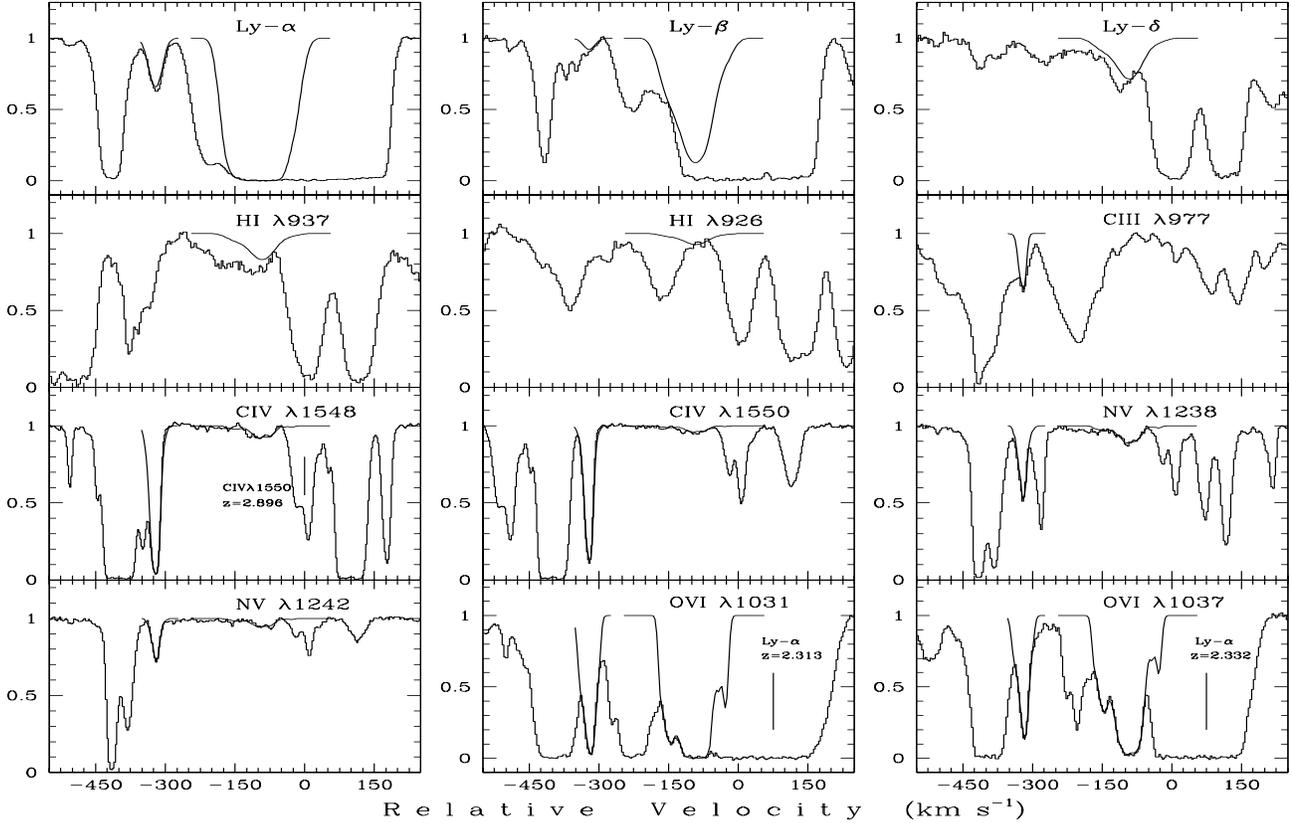,height=12cm,width=18cm}
\vspace{-0.7cm}
\caption[]{Same as Fig.~\ref{fg_1} but for
the \zabs = 2.898 
($v \simeq -317$ \kms) and \zabs = 2.901 ($v \simeq -85$ \kms) systems
towards \object{HE 2347--4243} (solid-line histograms).
The zero radial velocity is fixed at $z = 2.9021$.
Synthetic profiles corresponding to the
ionizing spectrum with a break at $\sim$4 Ryd (Fig.~\ref{fg_10})
are plotted by smooth curves. See text for details.
}
\label{fg_9}
\end{figure*}

This system (component 8 in Fechner et al. 2004)
exhibits strong lines of the doublets
\ion{C}{iv}, \ion{N}{v} and \ion{O}{vi} which allow us to estimate
accurately the covering factors and
column densities (Fig.~\ref{fg_9}). 
There is a weak hydrogen absorption Ly-$\alpha$, whereas the observed
intensity at the position of Ly-$\beta$
may be due to a blend with some forest line. 
However, these lines can be used to set 
a limit on the covering factor for \ion{H}{i}.
The \ion{C}{iii}$\lambda 977$ \AA\ line is also observed, albeit  blended.

The measured
column densities and covering factors for ions \ion{C}{iv}, 
\ion{N}{v} and \ion{O}{vi} are given in Table~\ref{tbl-4}.
The apparent
intensity at the position of Ly-$\beta$ constrains the covering 
factor and column density for \ion{H}{i}:
$C_{\scriptstyle \rm H\,{\scriptscriptstyle I}} < 0.6$, and
$N_{\scriptstyle \rm H\,{\scriptscriptstyle I}} < 2.3\times10^{13}$ \cm. 
The covering factor for \ion{C}{iii} can lie between
$C_{\scriptstyle \rm H\,{\scriptscriptstyle I}}$ and
$C_{\scriptstyle \rm C\,{\scriptscriptstyle IV}}$, i.e.
$0.60 < C_{\scriptstyle \rm C\,{\scriptscriptstyle III}} < 0.97$. 
The observed intensity at the expected position of the  
\ion{C}{iii} line allows us to estimate the
upper limits to its column density:
$N_{\scriptstyle \rm C\,{\scriptscriptstyle III}} < 9.0\times10^{12}$ \cm if 
$C_{\scriptstyle \rm C\,{\scriptscriptstyle III}} = 0.60$, and 
$N_{\scriptstyle \rm C\,{\scriptscriptstyle III}} < 4.7\times10^{12}$ \cm if 
$C_{\scriptstyle \rm C\,{\scriptscriptstyle III}} = 0.97$. 

Taking the ratio
$N_{\scriptstyle \rm C\,{\scriptscriptstyle IV}}/ 
N_{\scriptstyle \rm O\,{\scriptscriptstyle VI}} = 2.3$
and assuming solar 
relative abundance of carbon to oxygen, [C/O]~$\sim$0,  
we can estimate the ionization parameter $U$ for a given SED. 
The limits set above on 
$N_{\scriptstyle \rm C\,{\scriptscriptstyle III}}$  
reject unambiguously
any type of a power law spectrum as well as the AGN spectrum of 
Mathews \& Ferland (1987): for $U$ corresponding to [C/O]~$\sim$0
these spectra significantly overpredict 
$N_{\scriptstyle \rm C\,{\scriptscriptstyle III}}$.

In order to comply with the limits on  
$N_{\scriptstyle \rm C\,{\scriptscriptstyle III}}$
the ionizing spectrum should be quite hard at 
$1 < E < 4$ Ryd ($\alpha < 1.0$) and have a sharp
break at $E \sim$4 Ryd with the intensity depression by more than an order 
of magnitude. 
The spectrum of \object{HE 2347--4243} is
known up to the wavelength of about 350 \AA\ (2.6 Ryd)  
and it indeed shows a very hard
continuum in the EUV range with $\alpha$ = 0.56 
(Reimers et al. 1998; Fechner et al. 2004). 
Adopting this index, we can estimate
the depth of the intensity break at 4 Ryd: 
$J_{\rm A}/J_{\rm B} = 20$ (assuming 
$C_{\scriptstyle \rm C\,{\scriptscriptstyle III}} = 0.60$), and 
$J_{\rm A}/J_{\rm B} = 30$ (assuming 
$C_{\scriptstyle \rm C\,{\scriptscriptstyle III}} = 0.97$).  

With these EUV spectra, the
carbon and oxygen abundances in the \zabs = 2.898 system
can vary between 5-15 solar~-- 
depending on the adopted covering factor for \ion{H}{i} and the
depth of the continuum break at 4 Ryd, but 
nitrogen remains always underabundant: [N/O,C]~$ < 0$. Absolute
values of the abundances are uncertain  
not only due to unknown
covering factor of \ion{H}{i}, but also because of a possible overionization 
of \ion{H}{i} as a result of high temperature: calculations with CLOUDY
shows that the \ion{H}{i} fraction decreases more rapidly 
than the fractions of \ion{C}{iv} and
\ion{O}{vi} when temperature rises above the
thermal equilibrium level. 
On the other hand, time to reach the ionization equilibrium 
is longer for \ion{H}{i} than for other ions (Osterbrock 1974). It means that
calculations which use the  \ion{H}{i} fraction corresponding to 
the thermal equilibrium can significantly
underpredict the total hydrogen amount and 
deliver artificially boosted metallicity.
In principle, a neighbor system centered at $-415$ \kms 
(Fig.~\ref{fg_9}) may be considered as a support for this statement: 
although accurate calculations 
are prevented by line blending, an upper limit for the gas metallicity  
does not exceed 3 solar values. 
Unfortunately, it is impossible to distinguish between two
interpretations of absorption lines seen at 
$v < -325$ \kms: whether they are caused by the overlapping of clouds with
different metallicities or by clouds with similar metallicity 
but some of them not in thermal equilibrium.

\begin{table}[t]
\centering
\caption{
Physical parameters of the \zabs = 2.147 metal absorber
towards  \object{HE 1341--1020} derived by the MCI procedure
(uncertainties of the physical parameters are 10-15\%:
$U_0$~-- mean ionization parameter; $N_{\rm H}$~-- total hydrogen
column density; $\sigma_{\rm v}, \sigma_{\rm y}$~-- dispersions of
velocity and density distributions, respectively; 
$Z_{\rm X} = N_{\rm X}/N_{\rm H}$~-- abundances of the individual elements)
}
\label{tbl-3}
\begin{tabular}{lc}
\hline
\noalign{\smallskip}
Parameter & \\ 
\noalign{\smallskip}
\hline
\noalign{\smallskip}
$U_0$& 1.8E--2\\
$N_{\rm H}$, \cm& 1.8E20\\
$\sigma_{\rm v}$, \kms & 26.6\\
$\sigma_{\rm y}$& 0.65\\
$Z_{\rm C}$&2.2E--4 \\
$Z_{\rm N}$&3.9E--5 \\
$Z_{\rm O}$&7.0E--4 \\
$Z_{\rm Mg}$&4.4E--5\\
$Z_{\rm Al}$&8.5E--7 \\
$Z_{\rm Si}$&3.9E--5 \\
$Z_{\rm P}$&2.5E--7\\
$[Z_{\rm C}]^a$&$-0.03$ \\
$[Z_{\rm N}]$&$-0.18$ \\
$[Z_{\rm O}]$&$0.18$ \\
$[Z_{\rm Mg}]$&$0.10$ \\
$[Z_{\rm Al}]$&$-0.56$ \\
$[Z_{\rm Si}]$&$0.08$ \\
$[Z_{\rm P}]$&$0.10$ \\
$N$(H\,{\sc i}), \cm&$(8.6\pm1.5)$E16\\
$N$(He\,{\sc ii}), \cm&3.0E18$^b$ \\
$N$(C\,{\sc ii}), \cm&$(1.8\pm0.1)$E14 \\
$N$(C\,{\sc ii}$^\ast$), \cm&$(4.1\pm0.2)$E12 \\
$N$(C\,{\sc iii}), \cm&1.0E12$^b$ \\
$N$(C\,{\sc iv}), \cm&$(1.5\pm0.5)$E16 \\
$N$(N\,{\sc ii}), \cm&$(1.6\pm0.2)$E13 \\
$N$(N\,{\sc iii}), \cm&2.0E15$^b$ \\
$N$(N\,{\sc v}), \cm&$(7.1\pm1.5)$E14 \\
$N$(O\,{\sc vi}), \cm&$(3.0\pm0.5)$E15 \\
$N$(Mg\,{\sc ii}), \cm&$(1.5\pm0.2)$E13 \\
$N$(Al\,{\sc iii}), \cm&$(2.7\pm0.4)$E12 \\
$N$(Si\,{\sc ii}), \cm&$(8.4\pm0.8)$E12 \\
$N$(Si\,{\sc ii})$^\ast$, \cm&$\la 2.5$E11 \\
$N$(Si\,{\sc iii}), \cm&$(2.7\pm0.5)$E14 \\
$N$(Si\,{\sc iv}), \cm&$(7.0\pm1.5)$E14 \\
$N$(P\,{\sc v}), \cm&$(6.5\pm0.7)$E12 \\
$\langle T \rangle$, K & 1.0E4 \\
\noalign{\smallskip}
\hline
\noalign{\smallskip}
\multicolumn{2}{l}{$^a[Z_{\rm X}] = \log (N_{\rm X}/N_{\rm H}) -
\log (N_{\rm X}/N_{\rm H})_\odot$\ . }\\
\multicolumn{2}{l}{$^b$Calculated using the velocity and density distributions}\\
\multicolumn{2}{l}{\ \ derived from hydrogen and metal profiles.}
\end{tabular}
\end{table}

\subsubsection{System at \zabs = 2.901}

This system (centered at $v \simeq -85$ \kms\ in Fig.~\ref{fg_9},
component 10 in Fechner et al. 2004),
exhibits unblended weak lines of the doublets 
\ion{C}{iv} and \ion{N}{v}
along with strong saturated lines of \ion{O}{vi} partially 
blended in the wings. 
The line \ion{H}{i} Ly-$\alpha$
is saturated and black at the center which
means that the absorber completely covers the light source.
Assuming constant metallicity across the cloud, it is possible to
reconstruct the line profiles from their unblended parts.
The obtained column densities are given in Table~\ref{tbl-4}. 
A spectral feature
at $v \simeq -25$ \kms\ in the \ion{O}{vi} profile 
is due to the assumption of the 
constant metallicity and is uncertain since the corresponding parts in both
\ion{O}{vi} lines are blended. However, its input to the total column density
of \ion{O}{vi} is insignificant because 
$N_{\scriptstyle \rm O\, {\scriptscriptstyle VI}}$ 
is mostly determined
by the central portion of the \ion{O}{vi} lines which are clear.

The measured ratio 
$N_{\scriptstyle \rm O\,{\scriptscriptstyle VI}}/ 
N_{\scriptstyle \rm C\,{\scriptscriptstyle IV}} = 200$
is extremely high and points to the ionizing radiation
much stronger than the intergalactic: for the intergalactic absorbers 
with similar column densities of \ion{H}{i}
the ratio 
$N_{\scriptstyle \rm O\,{\scriptscriptstyle VI}}/ 
N_{\scriptstyle \rm C\,{\scriptscriptstyle IV}}$
is about 10-13 (Reimers et al. 2006). 
Thus, the system under study
is probably located close to the quasar. 

With the same ionizing spectrum as was determined for the 
\zabs = 2.898 system,
the ionization parameter $U$ can be estimated if 
the relative abundance of carbon to oxygen [C/O] is known.
It is interesting to note that the
condition [C/O]~$\sim$0 used to fix $U$ 
in the preceding sub-sections, cannot be fulfilled in the present case: 
starting at $U \sim$1 the ionization curves of 
\ion{O}{vi} and \ion{C}{iv} are almost
parallel giving the ratio of the ion fractions 
$\lg (\Upsilon_{\scriptstyle \rm O\,{\scriptscriptstyle VI}}/
\Upsilon_{\scriptstyle \rm C\,{\scriptscriptstyle IV}}) = 0.8$
and thus producing a stable overabundance of 
oxygen [O/C]~$\sim$0.2.  At lower $U$ ($\la 1$) the
overabundance of oxygen increases. 

From measurements in the Galactic and extragalactic
\ion{H}{ii} regions it is known that low-metallicity gas can be overenriched
in oxygen with the safe upper bound  
[O/C]~$ < 0.5$ (Henry et al. 2000; Nava et al. 2007). 
This gives $U > 0.2$ and metal content relative to solar
[C]~$< -1.0$, [O]~$< -0.5$, [N]~$< -1.2$. 
These relative abundances may indicate SNe~II explosions as a main
source of metal enrichment in the \zabs = 2.901 absorber.

Metal content of the absorbing gas redward to the \zabs = 2.901 
absorber 
(systems centered at $v = 0$ \kms\ and $v = 115$ \kms\ in Fig.~9)  
is uncertain due to
blending of both \ion{O}{vi} lines with forest absorptions. 
However, the column density of neutral
hydrogen can be estimated quite accurately
from the break at 912 \AA\ (rest frame) clearly seen
in the spectrum of \object{HE 2347-4243}: 
$N$(\ion{H}{i})~= $(1.6-2)\times10^{16}$ \cm. 
Depending on the ionization parameter of the gas and the spectral shape
of the incident ionizing radiation the corresponding
amount of \ion{He}{ii} ranges in the interval
$(2-5)\times10^{17}$ \cm and, thus, \ion{He}{ii} 
can soften the transmitted UV spectrum by 0.3-0.4 dex at $E > 4$ 
Ryd~\footnote{Smette et al. (2002) suggest for this system
$N$(\ion{He}{ii})~$\ga 2.4\times10^{18}$ \cm, which is a clear
overestimation: the extremely hard at $1 < E < 4$ Ryd spectrum
of \object{He 2347-4243} cannot provide 
$\eta = N$(\ion{He}{ii})/$N$(\ion{H}{i}) $> 100$ for 
any value of the break at 4 Ryd.}. 
Since relative location of the absorbers along the line of sight from
the quasar towards the observer is unknown, it cannot be
excluded that the absorber at \zabs = 2.898 follows in physical space after 
the absorbers considered here. This
means that the SED reconstructed from the \zabs = 2.898 
system may represent the quasar radiation softened partly by the
intervening gas seen at $-150$ \kms\ $< v < 150$ \kms.

\subsection{System at \zabs = 2.352 towards \object{Q0329--385} }

The spectrum of the quasar \object{Q 0329--385} 
was obtained with the UVES/VLT in the framework of the ESO Large
Program `QSO Absorption Line Systems' (ID No.166.A-0106).
Data reduction was performed by B.~Aracil. 
The system at \zabs = 2.352
was firstly described in Bergeron et al. (2002) in their study of intergalactic 
\ion{O}{vi} absorbers and recently by Schaye et al. (2007).

\begin{table}[t!]
\centering
\caption{Column densities and covering factors for the associated systems
at \zabs = 2.898 and 2.901 towards \object{HE 2347--4243}
shown in Fig.~\ref{fg_9} }
\label{tbl-4}
\begin{tabular}{ccccc}
\hline
\hline
\noalign{\smallskip}
 & \multicolumn{2}{c}{$z = 2.898$} & \multicolumn{2}{c}{$z = 2.901$} \\[-2pt]
Ion & $N$, cm$^{-2}$ & $C$  & $N$, cm$^{-2}$ & $C$ \\[-2pt]
\noalign{\smallskip}
\hline 
\noalign{\medskip}
H\,{\sc i} & $2.3$E13 &  0.6 & $(7.6\pm1.0)$E14 & 1.0 \\
           & $1.1$E13 &  0.97  \\
C\,{\sc iii} & $9.0$E12 &  0.6  \\
             & $4.8$E12 &  0.97  \\
C\,{\sc iv} & $(1.2\pm0.1)$E14 & 0.97 & $(6.7\pm1.0)$E12 & 1.0 \\
N\,{\sc v} & $(2.6\pm0.2)$E13 & 0.97 & $(1.5\pm0.3)$E13 & 1.0 \\
O\,{\sc vi} & $(2.8\pm0.2)$E14 & 0.985 & $(1.3\pm0.2)$E15$^a$ & 1.0 \\
\noalign{\smallskip}
\hline
\noalign{\smallskip}
\multicolumn{5}{l}{$^a$Both \ion{O}{vi}$\lambda\lambda 1031, 1037$ \AA\ lines
are blended in the red wings}\\ 
\multicolumn{5}{l}{which are restored from v-d distributions obtained from}\\
\multicolumn{5}{l}{other metal lines and assuming constant metallicity within}\\
\multicolumn{5}{l}{the absorber. The column density is set by the portion}\\
\multicolumn{5}{l}{$-60 < v < 40$ \kms which is unblended.}
\end{tabular}
\end{table}

The redshift of the quasar
\object{Q 0329--385} is \zem = 2.435 (\ion{H}{i} Ly-$\alpha$)  which means
that the \zabs = 2.352
system is detached from the QSO by $\sim$7400 \kms. 
There are clear lines of the doublet \ion{C}{iv}, whereas
doublets \ion{N}{v} and \ion{O}{vi} are blended each in one component 
(Fig.~\ref{fg_11}). 
The relative intensities of both \ion{C}{iv} lines do not show any evidence of
the incomplete coverage of the light source. 
The system is very much alike to that at \zabs = 2.898 towards 
\object{HE 2347-- 4243} (Sect.~\ref{Ss-5}): strong lines of
highly ionized metals combined with a weak \ion{H}{i} line.
 
The measured column densities are given in Table~\ref{tbl-5}. 
The present system shows the column density ratio
$N_{\scriptstyle \rm O\,{\scriptscriptstyle VI}}/ 
N_{\scriptstyle \rm C\,{\scriptscriptstyle IV}} = 3.7$
which is only slightly higher than 
$N_{\scriptstyle \rm O\,{\scriptscriptstyle VI}}/ 
N_{\scriptstyle \rm C\,{\scriptscriptstyle IV}} = 2.3$
of the \zabs = 2.898 system, but it contains much more nitrogen: 
$N_{\scriptstyle \rm N\,{\scriptscriptstyle V}}/ 
N_{\scriptstyle \rm C\,{\scriptscriptstyle IV}} = 1.34$
compared to  
$N_{\scriptstyle \rm N\,{\scriptscriptstyle V}}/ 
N_{\scriptstyle \rm C\,{\scriptscriptstyle IV}} = 0.21$
at \zabs = 2.898.

The apparent profile of 
Ly-$\alpha$ is inconsistent with the
assumption of constant metallicity throughout the absorber (synthetic profiles
are shown by the smooth curves in Fig.~\ref{fg_11})\footnote{The velocity 
shift between the dominant
component in \ion{H}{i} and metal lines was noticed for 
the \zabs = 2.352 system also by Shaye et al. (2007).}. 
The \ion{H}{i} Ly-$\beta$ line is blended and can be employed merely to
set a lower limit for the covering factor and, correspondingly, 
an upper limit for $N$(\ion{H}{i}). 
Since the \ion{C}{iii}$\lambda 977$ \AA\ line 
is blended with a strong line \ion{C}{iv}$\lambda 1548$ \AA\ from the
\zabs = 1.115 system, the only way to fix the ionization parameter 
is to use the condition [C/O]~$\sim$0.

\begin{figure}[t]
\vspace{0.0cm}
\hspace{-0.2cm}\psfig{figure=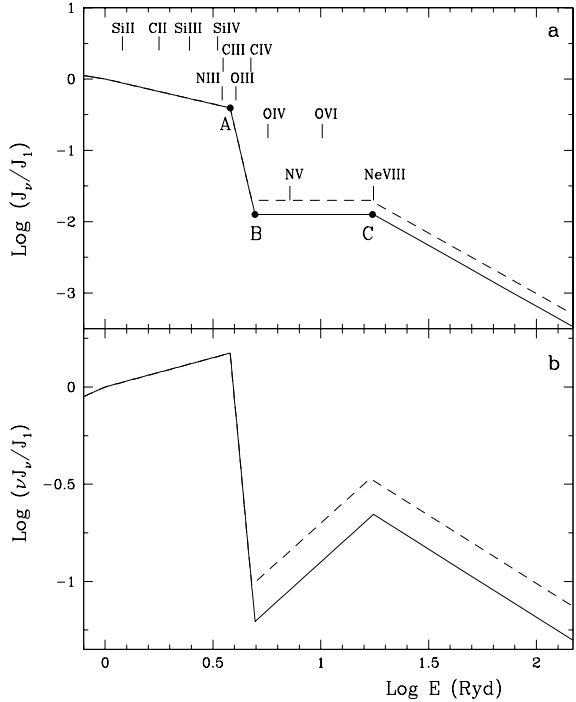,height=10cm,width=9.7cm}
\vspace{-1.0cm}
\caption[]{SED restored from the \zabs = 2.898 system towards 
the quasar \object{HE 2347--4243} (\zem = 2.902).
The dashed line represents the ionizing spectrum 
corresponding to the covering factor 
$C_{\scriptstyle \rm C\,{\scriptscriptstyle III}} = 0.6$, 
solid line~-- to
$C_{\scriptstyle \rm C\,{\scriptscriptstyle III}} = 0.97$ 
(see text for details).
}
\label{fg_10}
\end{figure}

\begin{figure*}[t]
\vspace{0.0cm}
\hspace{-0.2cm}\psfig{figure=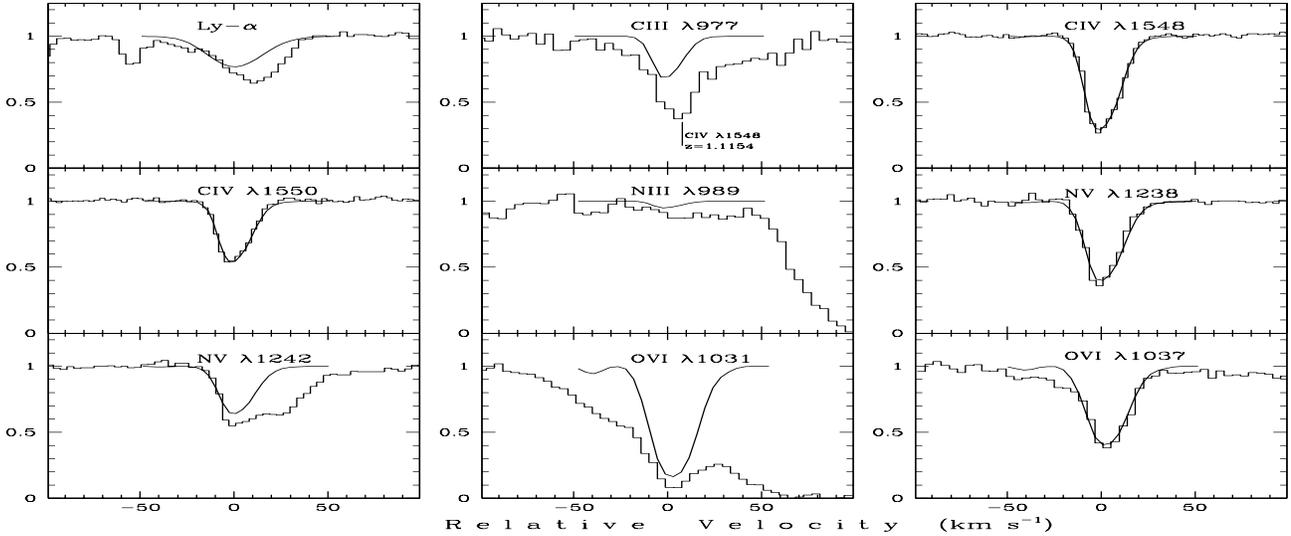,height=9cm,width=18cm}
\vspace{-1.0cm}
\caption[]{Same as Fig.~\ref{fg_1} but for
the \zabs = 2.352 system
towards \object{Q 0329--385} (solid-line histograms).
The zero radial velocity is fixed at $z = 2.35205$.
Synthetic profiles are plotted by smooth curves.
}
\label{fg_11}
\end{figure*}

With this restriction, any reasonable spectral index $\alpha$ 
of a pure power law SED ($\alpha = 1.0-1.8$) leads to
the C and O abundances of 10-20 solar and to the
nitrogen abundance of 20-40 solar.
These values seem to be definitely too high, even if one assumes 
that hydrogen is overheated 
(in fact, there are no signs of possible overheating in this system). 
On the other hand,
for the ionizing spectra with a break at 4 Ryd the 
condition [C/O]~$\sim$0 is fulfilled at higher $U$ than
for the power law spectra, resulting in a lower fraction of neutral hydrogen 
and, hence,  a higher amount of total hydrogen 
whereas the fractions of \ion{C}{iv} and \ion{O}{vi} do not change. 
Thus, the metal abundances decrease: the softer the spectrum at 
$E > 4$ Ryd the lower metallicity. For example, the spectrum restored
for the \zabs = 2.898 system (Fig.~\ref{fg_10}, solid line) 
gives [C]~= [O]~= 0.5 and [N] = 0.9. 
The accurate value of the continuum depression at 4 Ryd cannot be estimated 
since the real metallicity of the gas cloud at
\zabs = 2.352 is of course unknown. 
However, we can conclude that the SED with a break at 4 Ryd
is clearly preferable for this system. 
Another conclusion is a revealed significant relative 
overabundance of nitrogen, [N/C]~$\sim$0.3-0.4, 
which was constantly reproduced for any tried ionizing spectrum.

\begin{table}[t!]
\centering
\caption{Column densities and covering factors for the associated system
at \zabs = 2.352 towards \object{Q 0329--385}
shown in Fig.~\ref{fg_11} }
\label{tbl-5}
\begin{tabular}{ccc}
\hline
\hline
\noalign{\smallskip}
Ion & $N$, cm$^{-2}$ & $C$ \\
\noalign{\smallskip}
\hline 
\noalign{\medskip}
H\,{\sc i} & $(1.3-1.4)$E13$^a$ &  0.6 \\
           & $(7-8)$E12$^a$ &  1.0  \\
C\,{\sc iii} & $3.6$E12$^b$ & 1.0 \\
C\,{\sc iv} & $(3.2\pm0.1)$E13 & 1.0 \\
N\,{\sc iii} & $2.9$E12$^b$ & 1.0 \\
N\,{\sc v} & $(4.0\pm0.2)$E13 & 1.0 \\
O\,{\sc vi} & $(1.2\pm0.1)$E14 & 1.0 \\
\noalign{\smallskip}
\hline
\noalign{\smallskip}
\multicolumn{3}{l}{$^a$The red wing of the apparent profile inconsistent with}\\
\multicolumn{3}{l}{the assumption of constant metallicity across the}\\
\multicolumn{3}{l}{absorber. The synthetic profile is restored on base of}\\
\multicolumn{3}{l}{velocity-density (v-d) distributions obtained from metal}\\
\multicolumn{3}{l}{lines and assuming homogeneous metallicity.}\\
\multicolumn{3}{l}{$^b$Estimated from v-d distributions obtained from the}\\
\multicolumn{3}{l}{metal lines \ion{C}{iv}, \ion{N}{v}, \ion{O}{vi}
and SED shown in Fig.~\ref{fg_10}. }
\end{tabular}
\end{table}

\section{Discussion}

\subsection{The origin of the break at 4 Ryd}

All NAL systems considered above as well as NAL systems towards 
\object{HE 0141--3932} (Reimers et al. 2005)
are best described with ionizing spectra having a break at
$E \approx 4$ Ryd. It seems natural to relate this break to 
the Lyman continuum absorption in \ion{He}{ii}.
Then the question arises where does this absorption occur?

Firstly, this absorption can be produced by the NAL systems themselves~-- 
by those with $N$(\ion{H}{i})~$> 10^{16}$ \cm. 
This possibility was suggested in 
Reimers et al. (1997) and then discussed also in Smette et al. (2002) and
Shull et al. (2004). 
Among considered absorbers, to this group belong the system
at \zabs = 1.7103 towards \object{HE 0141--3932} 
(Reimers et al. 2005), and systems at \zabs = 2.901 
(\object{HE 2347--4243}) and \zabs = 2.147 (\object{HE 0141--3932})
from the present work.
The \zabs = 2.147 system with $N$(\ion{H}{i})~$\approx 10^{17}$ \cm\
and $N$(\ion{He}{ii})~$\approx 3\times10^{18}$ \cm\
can noticeable soften the incident radiation due to internal
absorption in the \ion{He}{ii} Lyman continuum,
but the metal lines detected show that the incident radiation
should already have a significant break at $E = 4$ Ryd (Sect.~2.2.1).
The \zabs = 1.7103 and \zabs = 2.901 systems have 
$N$(\ion{He}{ii})~$< 5\times10^{17}$ \cm\ and also cannot
account for the whole depth of the break at 4 Ryd in the
reconstructed spectra. Besides, lines of sight towards
\object{J 2233--606} and \object{Q 0329--385} with the associated
systems at \zabs = 2.198 (Sect.~2.1.1) and \zabs = 2.352 (Sect.~2.4),
both strongly favoring the ionizing spectrum broken at 4 Ryd,
do not show any absorption with
$N$(\ion{H}{i})~$\sim 10^{16}$ \cm\ in the range $z_{\rm em} - z < 0.3$.
Thus, in general the \zabs~$\approx$ \zem systems
can hardly be the main source of the observed \ion{He}{ii} Lyman 
continuum absorption. 

More probably, the bulk of the 4 Ryd discontinuity is caused by 
the \ion{He}{ii} absorption in quasar winds
which show a common occurrence both in quasars and AGNs.
The accretion disk winds are supposed to be accelerated by the 
combination of radiation-pressure and  magneto-centrifugal forces
and comprise dense gas near the equator seen as UV broad absorption lines
(BAL) and highly ionized shielding gas which protects the inner wind
regions from overionization by X-rays from the central source
(Murray et al. 1995; Proga et al. 2000; Everett 2005). 
All quasars considered here do not belong to BAL quasars 
(only \object{HE~0141--3932} can be classified as mini-BAL).
The natural question is whether their spectra contain any features which
allow to conclude that the wind absorption is indeed present and that
the column density of \ion{He}{ii} in the outflowing gas is of order
$\sim$$10^{18}$ \cm\ (the depth of the 4 Ryd break $\sim$1 dex).

Model calculations with CLOUDY show that for any incident ionizing spectrum
the ratio $\eta$ = $N$(\ion{He}{ii})/$N$(\ion{H}{i}) firstly increases with
the rise of the ionization parameter $U$, but attains the constant value  
when $U$ becomes large (Fig.~19 in Agafonova et al. 2007).
For power law spectra this ratio equals $20-40$ depending on the slope
of the spectrum (the steeper spectrum the larger $\eta$) 
and remains almost constant at $U > 0.1$. 
Thus, $N$(\ion{He}{ii}) of $\sim$$10^{18}$ \cm\ corresponds to $N$(\ion{H}{i})
of a few times $10^{16}$ \cm. 
The ionization parameter $U$ can be constrained if
we assume the total column density of hydrogen 
$N$(\ion{H})~$\sim$$10^{22}-10^{23}$ \cm\ 
which is estimated from X-ray observations of quasars
(Murray et al. 1995; Piconcelli et al. 2005). 
This gives $U \sim$1-5. 
In this $U$ range, among ions observable in the UV the 
highest fractions exhibit  \ion{O}{vi} (1/30~-- 1/100 of all oxygen) and
\ion{Ne}{vi}, \ion{Ne}{vii}, \ion{Ne}{viii} 
(1/4~--1/10 of all neon) whereas the ionization fraction 
of \ion{C}{iv} is negligible. 
Gas with such parameters 
is expected to leave in a UV quasar spectrum its footprints 
seen as    
\ion{H}{i} Ly-$\alpha$ $\lambda1215$ \AA\ absorption, 
\ion{H}{i} break at  $\lambda \leq 912$ \AA\ and~-- 
in case of high metallicity~-- 
\ion{O}{vi} $\lambda\lambda1032, 1037$ \AA, 
\ion{Ne}{vi} $\lambda\lambda401, 430$ \AA\, 
\ion{Ne}{vii} $\lambda465$ \AA, and
\ion{Ne}{viii} $\lambda\lambda770, 780$ \AA\ absorption features. 
Since the gas velocity range in the accretion disk
winds exceeds 10000 \kms, 
these absorption features should look like wide and shallow troughs, and 
a break at the hydrogen ionization edge~-- as a slow
continuous depression towards higher frequencies.
It is to note that in the given $U$ range the oxygen 
exists mostly in the form of \ion{O}{vii} (almost 90\% of all oxygen) and
\ion{O}{viii} ($\sim$10\% of oxygen), i.e. is best accessible in the X-ray
region.

\begin{figure*}[t]
\vspace{0.0cm}
\hspace{-0.2cm}\psfig{figure=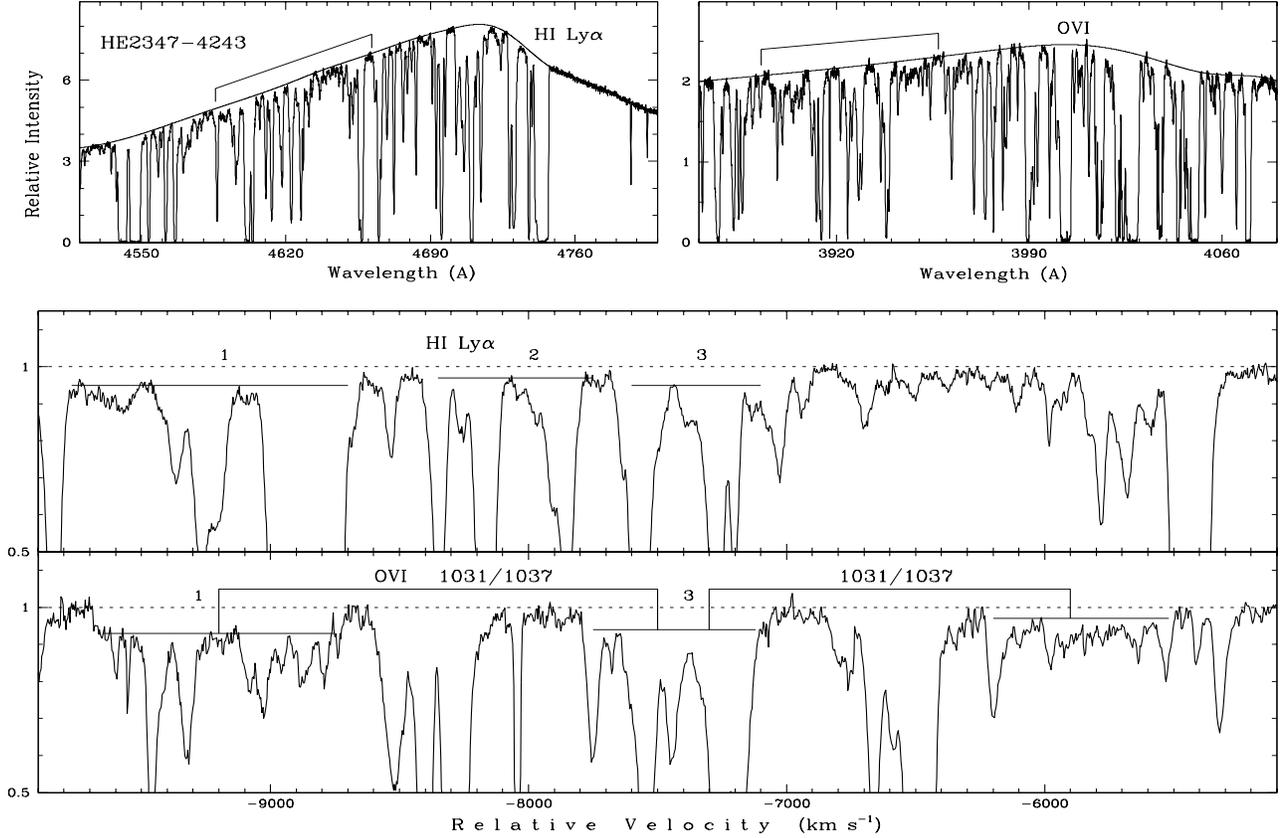,height=15cm,width=18cm}
\vspace{-3.3cm}
\caption[]{Broad and shallow absorptions detected in
the spectrum of \object{HE 2347--4243}.
The zero radial velocity is fixed at $z = 2.902$.
For other details see capture to Fig.~\ref{fg_8}.
}
\label{fg_12}
\end{figure*}

\begin{figure*}[t]
\vspace{0.0cm}
\hspace{-0.2cm}\psfig{figure=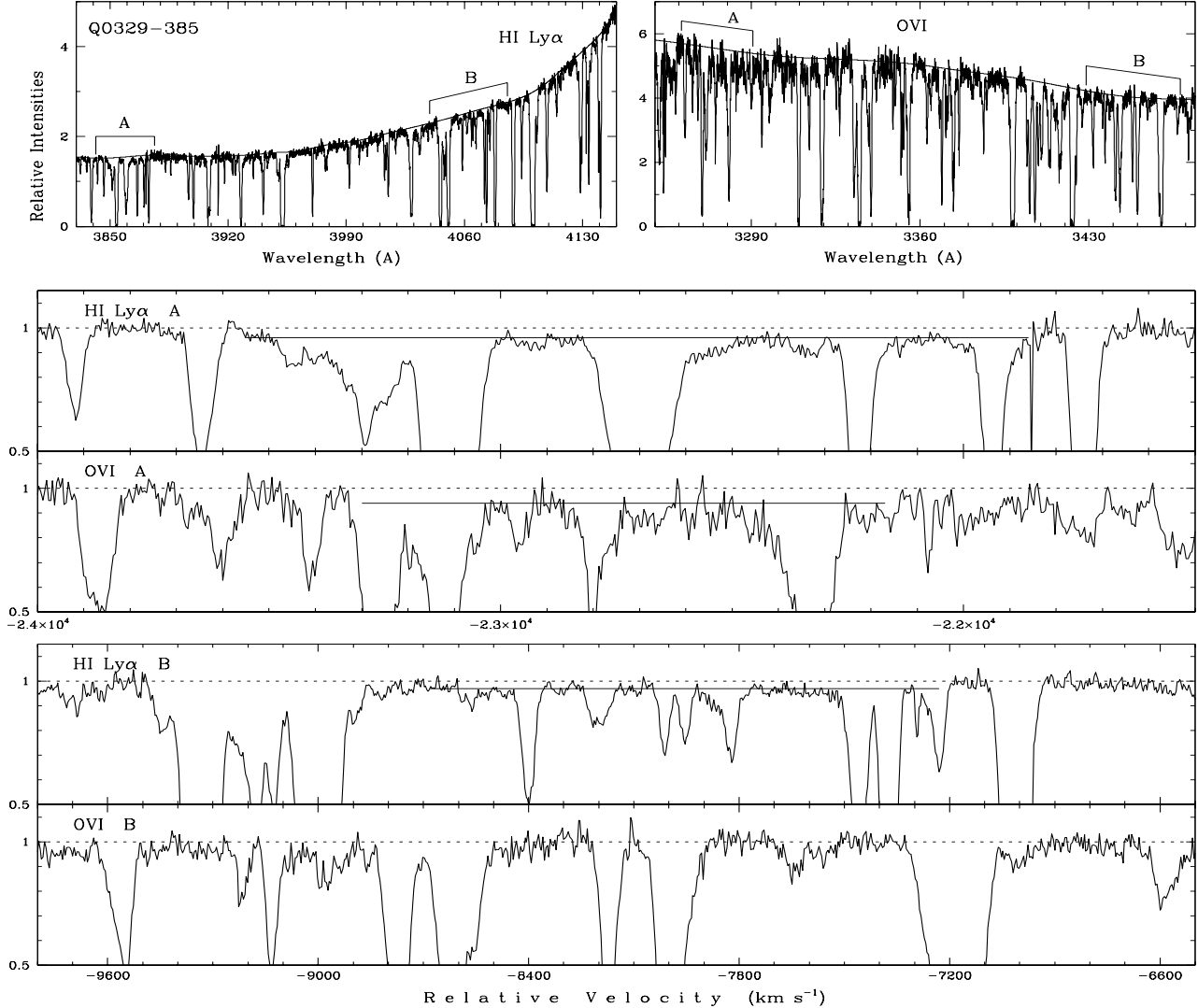,height=15cm,width=18cm}
\vspace{-0.3cm}
\caption[]{Broad and shallow absorptions detected in
the spectrum of \object{Q 0329--385}.
The zero radial velocity is fixed at $z = 2.435$.
For other details see capture to Fig.~\ref{fg_8}.
}
\label{fg_13}
\end{figure*}

We do find the expected features in the
quasar spectra considered in the present work. 
The depression at $\lambda \leq 912$ \AA\ (rest-frame) 
is clearly seen in the flux-calibrated spectrum of \object{J 2233--606} 
shown in Fig.~1 in Sealy et al. (1998) and its depth 
($\tau^{\rm H\,{\scriptscriptstyle I}}_c \sim$0.3)
corresponds to 
$N$(\ion{H}{i})~$\sim$$4\times10^{16}$ \cm. 
Depression occurring in the flux-calibrated
spectrum of \object{HE 2347--4243} (Fig.~3 in Fechner et al. 2004)
accounts for $N$(\ion{H}{i}) from 3 to $7\times10^{16}$ \cm. 
For \object{HE 1341--1020}, the corresponding wavelength range is
beyond the observable frame, 
whereas for \object{Q 0329--325} the flux-calibrated spectrum is not available. 
It should be noted that the flux-calibration is the mandatory requirement 
to detect this slow rolling continuum depression because
otherwise it is disguised by instrumental effects.
Further on, shallow troughs of hydrogen Ly-$\alpha$ 
and \ion{O}{vi} absorptions extending over thousands \kms\
are present in spectra of 
\object{HE 0141--1020} (Fig.~\ref{fg_8}, marked by horizontal lines),  
\object{HE 2347--4243} (Fig.~\ref{fg_12}) and
\object{Q 0329--385} (Fig.~\ref{fg_13}). 
In spite of very low optical depth of \ion{H}{i} Ly-$\alpha$ 
($\tau^{\rm H\,{\scriptscriptstyle I}}_\alpha \sim$0.03), 
the corresponding
portions can be clearly distinguished in the spectra 
due to high signal-to-noise ratio and high spectral
resolution which allows us to trace `clear continuum windows' between
absorption features over a wide spectral range. 
We also checked whether the detected troughs could be
due to mismatch of echelle orders. The answer is no: in the spectral
range considered the orders are small ($\sim$300 \kms), and a depression
of the width of $\sim$1000s \kms covers several subsequent orders.

The optical depth  of an absorption line arising in a gas flow moving 
with a large velocity gradient is given by 
(Sobolev 1947, 1960; Castor et al. 1975):
\begin{equation}
\tau = 2.654\times10^{-15}\,f\lambda\,n/ \Bigl | 
\frac{dv}{ds} \Bigr | \ ,
\label{E8}
\end{equation}
where $f$ is the oscillator strength for absorption, $\lambda$ the central
wavelength (in \AA), $n$ the number density (in \cmm) of atoms/ions in
the ground state, $v$ the wind speed (in \kms), 
and $s$ the distance (in cm) along the line of sight.  
The population of the upper excited level is ignored. 

For \ion{H}{i} Ly-$\alpha$ and \ion{O}{vi} $\lambda 1031$ 
(as an example of a heavy element) 
the corresponding optical depths are 
$$
\tau^{\rm H\,{\scriptscriptstyle I}}_\alpha = 
1.34\times10^{-12}\,n_{\rm H}\ \Upsilon_{\rm H\,{\scriptscriptstyle I}}/
\Bigl |\frac{dv}{ds} \Bigr | \, ,
$$
and
$$
\tau^{\rm O\,{\scriptscriptstyle VI}}_{1031} = 
3.63\times10^{-13}\, 
n_{\rm H}\, Z_{\rm O}\,
{\Upsilon_{\rm O\,{\scriptscriptstyle VI}}}/
\Bigl |\frac{dv}{ds} \Bigr |\, , $$
where $Z_{\rm O}$ is the oxygen abundance and 
$\Upsilon_{\rm H\,{\scriptscriptstyle I}}$, 
$\Upsilon_{\rm O\,{\scriptscriptstyle VI}}$ are the ionization fractions
of \ion{H}{i} and \ion{O}{vi}, respectively.

The viable estimation for the velocity acceleration 
$dv/ds$ at $v > 1000$ \kms\ is $10^{-7} - 10^{-8}$ km~s$^{-1}$~cm$^{-1}$ 
(e.g., Murray et al. 1995) with lower acceleration corresponding to higher
velocities.
Thus, to produce the optical depth of $\sim$0.03 the
number density of neutral hydrogen should be 
$n_{\rm H\,{\scriptscriptstyle I}} \sim$(2-20)$\times10^2$ \cmm. 
For an ionization parameter $U \sim$1-5 this gives
the gas number density of 
$n_{\rm H} = 10^8 - 10^9$ \cmm. 
Absorption in a gas with lower 
density will be indistinguishable from
noise fluctuations in the quasar continuum windows whereas
more dense parts of the disk wind will produce pronounced
absorption which can be seen not only in \ion{H}{i}, \ion{O}{vi}, and
\ion{Ne}{viii}
but -- due to lower $U$ -- in such
ions as \ion{C}{iv}, \ion{N}{v} or even \ion{Si}{iv} as well. 
Large density fluctuations are expected in the disk wind
due to flow stratification and due to the presence of
various instabilities which give rise to shocks and dense shells as
they propagate from the wind base outward and 
grow rapidly to non-linear regime (Owocki et al. 1988; Proga et al. 2000).
Thus, in general, the wind absorption is represented by the
structure stretching over 1000s \kms\ and consisting of the
low-contrast continuum depression intermitted by parts without any absorption
and by deeper absorption
features caused by both density and velocity fluctuations\footnote{cf. 
Levshakov \& Kegel (1998), where a `line-like' structure caused by
velocity fluctuations superposed on the general
Hubble flow was calculated in the approximation of the constant gas density
and neglecting the coupling of the velocity field with the radiation field.}. 
In the quasar spectra considered here the broad
and shallow absorption features can be detected at radial velocities up to
20000 \kms\ from the quasar's systemic redshifts which corresponds to 
$\Delta z \sim$0.2 at $2 < z < 3$. 
Thus, the superposition of absorption lines 
from the intervening Ly-$\alpha$ forest clouds is inevitable.
Absorptions of \ion{H}{i} and
\ion{O}{vi} in Fig.~\ref{fg_8}, \ref{fg_12}, 
\ref{fg_13} show all elements of this picture\footnote{Similar 
complex  absorption-line system consisting of shallow
and broad ($FWHM \sim$700 \kms) and narrow ($FWHM < 20$ \kms) absorption
lines displaced by $\simeq$14000 \kms blueward of the quasar systemic
velocity was detected at $z = 3.021$ towards \object{CTQ 325} (Levshakov et al.
2004).}.

Another 
important question is to what extent the quasar's outcoming radiation is 
affected by the \ion{He}{ii} Lyman continuum absorption. It seems quite
probable that outflow zones which produce BALs are opaque in \ion{He}{ii}
continuum. X-ray observations of BAL quasars show that X-ray absorbing gas  
has 
$N$(\ion{H})~= $10^{22}-10^{24}$ \cm\ (Gallagher et al. 2002). 
The BAL systems
are mostly identified by strong \ion{C}{iv} absorption which supposes
that the ionization parameter $U$ is less than 1 for a power law
incident continuum (or that incident continuum is soft at E > 4 Ryd). 
This gives the column density of neutral hydrogen 
$N$(\ion{H})$\sim$$10^{17}$ \cm\ and of single ionized helium 
$N$(\ion{He}{ii}) $\sim$ few times of $10^{18}$ \cm. Thus, the
quasar spectrum transmitted through the BAL-producing region should have   
a pronounced break (more than 1 dex) at 4 Ryd. However, BAL quasars
comprise only about 15\% of quasars population (Reichard et al. 2003)
what means~-- if one considers the BAL effect as being due to relative
orientation of the disk and the line of sight~-- that BAL gas 
affects an insignificant part of the outcoming radiation.

As mentioned above, $N$(\ion{He}{ii}) of $\sim$$10^{18}$ \cm\ corresponds to
$N$(\ion{H}{i}) of few times $10^{16}$ \cm.
This amount of neutral hydrogen gives origin to the continuum
depression by $\sim$15-30\% starting somewhere above 912 \AA\ in the
rest-frame -- the shielding gas can not only outflow but infall as well
(Murray et al. 1995; Proga et al. 2000) shifting 
the depression to the red. 
For non-BAL QSOs, the presence of such depression would
indicate that there is enough \ion{He}{ii} to produce a
noticeable intensity break at energies above 4 Ryd in the outcoming radiation.
It seems natural to assume that the depth of this break
depends on the angle to the rotational axis of the disk.
A statistical study of the 
hydrogen Lyman continuum depression in the flux-calibrated
high-resolution quasar spectra is expected to deliver a comprehensive
information about the covering factor of the \ion{He}{ii} opaque gas. This
study is still to be done. However, there are several hints that this
covering factor may indeed be significant. Continuum depression 
starting roughly at $\sim$930 \AA\ (rest-frame) is clearly seen in HST
spectra of non-BAL quasars shown in Binette et al. (2005) 
who state that such quasars (dubbed by them as
'class A' quasars) comprise about 60\% of their sample (106 species). 
The estimates of the corresponding column densities of \ion{H}{i} are not very
certain due
to the low quality of spectra, but nevertheless they vary 
between 2 and
$6\times10^{16}$ \cm~-- well within the predicted boundaries. 
A good example represents a
HST spectrum of \object{HS 1103+6416} 
shown in Fig.~1 in K\"ohler et al. (1999) -- 
continuum depression starting at $\sim$950 \AA\,
allows an accurate estimate of 
$N$(\ion{H}{i})~$\simeq 2.5\times10^{16}$ \cm. 
 
It is to note that so-called `composite' spectra obtained by
co-adding of many individual QSO spectra and shown in Fig.~5 in
Telfer et al. (2002) and in Fig.~3 in Scott et al. (2004) are
of no use in the present context since
fine features present in an individual spectrum vanish or
are smeared out in the co-added one.
Moreover, before co-adding the individual spectra were
`statistically corrected' for the Ly-$\alpha$ forest absorption
using the general distribution of the Ly-$\alpha$ clouds
over $N$(\ion{H}{i}) and $z$. 
This operation is quite subjective because 
statistical properties of the  Ly-$\alpha$ clouds distribution
vary significantly from sightline to sightline.
However, even in the composite spectra some feature are present
which may be attributed to the quasar wind:
these are
a trough-like structure between 700 and 900 \AA\ 
in Fig.~5 (Telfer et al. 2002), and absorption features blueward of
\ion{O}{vi} and \ion{Ne}{viii} emission lines again in
Fig.~5 (Telfer et al.) and in Fig.~3 in Scott et al. (2004).  
The former may be caused by the \ion{H}{i} Lyman continuum absorption,
and the latter~-- by the resonance absorptions in the 
\ion{O}{vi} and \ion{Ne}{viii} lines.

An additional support for possible \ion{He}{ii} opacity of quasar winds
comes from X-ray observations which detect in $\sim$50\% of quasars
so-called `warm absorbers' -- broad absorption feature due to the blend
of various ionization edges, among them \ion{O}{vii} and \ion{O}{viii}
(Piconcelli et al. 2005). There are evidences
that these absorbers have  
$N$(\ion{H})~$\sim 10^{22}-10^{23}$ \cm\ and
the ionization parameters $U \la 5$ 
(Krongold et al. 2003; Schartel et al. 2005;
Steenbrugge et al. 2005; Jimenez-Bail\'on et al. 2007).
This leads again to $N$(\ion{He}{ii})~$\sim$$10^{18}$ \cm.

\begin{figure}[t]
\vspace{0.0cm}
\hspace{-0.2cm}\psfig{figure=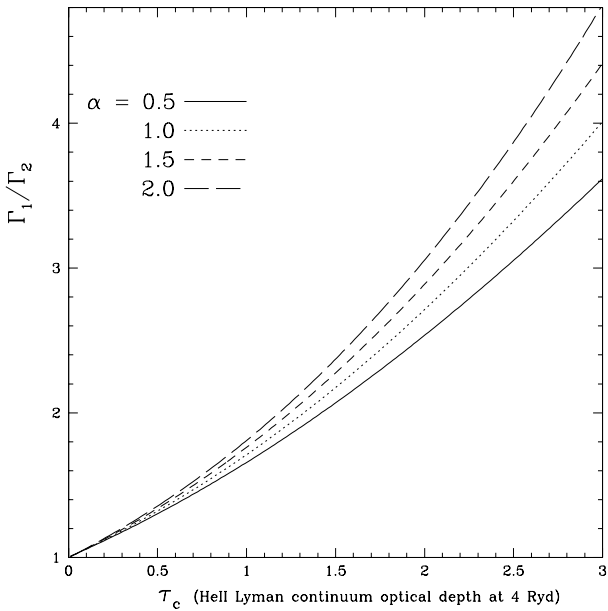,height=10cm,width=9.7cm}
\vspace{-1.2cm}
\caption[]{The ratio of the photoionization rates $\Gamma_1$ (a pure
power law spectrum $F_\nu \propto \nu^{-\alpha}$) to 
$\Gamma_2$ (a power law spectrum with a break
at 4 Ryd) as a function of 
$\tau^{\rm He\,{\scriptscriptstyle II}}_c$. 
}
\label{fg_14}
\end{figure}

To summarize, it seems quite likely that the outcoming QSO radiation
in general is not
a pure power law but has an intensity  break at $E \geq 4$ Ryd due to
the \ion{He}{ii} Lyman continuum absorption in the quasar disk wind. 
This can affect  
the rate of the \ion{He}{ii} reionization in the intergalactic clouds.
Recent observations of
the \ion{H}{i}/\ion{He}{ii} Ly$\alpha$ forest towards
\object{HE 2347--4342} (Shull et al. 2004; Zheng et al. 2004)
and \object{HS 1700+6416} 
(Fechner et al. 2006; Fechner \& Reimers 2007)
revealed significant fluctuations of 
$\eta$ which suggest a variable softness parameter
$S = J_{\rm 1 Ryd}/J_{\rm 4 Ryd}$
of the metagalactic radiation field.
In this respect the revealed \ion{He}{ii} opacity of the
quasar wind which changes depending on angle to the rotational axis of the disk and 
by no doubt differs also from quasar to quasar can 
be responsible for the local fluctuations of $S$. 
For a pure power law ionizing spectrum, $F_\nu \propto \nu^{-\alpha}$,
the photoionization rate of \ion{He}{ii} is 
$\Gamma_1 \propto 1/(\alpha+3)$. 
For a spectrum with a break at 4 Ryd,
$\tau^{\rm He\,{\scriptscriptstyle II}}_c > 0$,
the rate, $\Gamma_2$, becomes slower : 
$$
\Gamma_2 \propto \int\limits^\infty_1\frac{\exp(-
\tau^{\rm He\,{\scriptscriptstyle II}}_c/x^3)}{x^{\alpha+4}}dx,\,\,\,   
{\rm with}\,\,\, x = \frac{E}{\rm 4\, Ryd}\ .
$$
The ratio $\Gamma_1/\Gamma_2$ as a function of
$\tau^{\rm He\,{\scriptscriptstyle II}}_c$ is shown in Fig.~\ref{fg_14}.

\subsection{Element abundances in quasar associated systems}

The properties of the NAL systems themselves are rather a by-product of the
present study. However, there are some interesting results obtained which
are worth to be mentioned.

Metal content as well as the relative elemental abundances (N/C, 
$\alpha$-element/C, etc.) 
measured from the absorption systems are usually used 
to conclude about the enrichment mechanisms of the absorbing gas.
Concerning the quasar-related absorption systems
analyzed in the present work, the measured ratio N/C  
show that these mechanisms can be different for gas in the vicinity 
of the central engine and in the host galaxy. 
For instance, the systems at \zabs = 2.198 
(\object{J 2233--606}), \zabs = 2.107 (\object{HE 1341--1020})
and \zabs = 2.352 (\object{Q 0329--385}), 
all with systemic velocities of thousands of \kms
and, thus, parts of the outflows from the central region, 
reveal oversolar metallicity along with the relative overabundance 
of nitrogen as compared to solar values, [N/C]~$ > 0$. 
On the other hand, the systems at \zabs = 2.147 
(\object{HE 1341--1020}) and  \zabs = 2.898 (\object{HE 2347--4243}), 
with the systemic velocities of $\sim$100-200 \kms\ 
and belonging probably to the gas
in the host galaxy (cf. Nagao et al. 2006a), 
demonstrate similar oversolar metal content,
but underabundant nitrogen, [N/C]~$ < 0$.
Overabundance of nitrogen, a common finding also in the analysis of quasar broad
emission lines (e.g., Dietrich et al. 2003; Nagao et al. 2006b), 
is supposed to be a consequence 
of rapid stellar evolution of quasar gas 
which leads to scaling of
N/O ratio with metallicity (Hamann\& Ferland, 1999). 
However, different ratios [N/C]
at similar metallicity measured in the circumnuclear gas and in the host galaxy 
indicates different enrichment mechanisms which may be due to
variations in the initial mass function.
Note that underabundance of nitrogen not complying
with theoretical predictions of [N/C] and [Fe/C] was measured in the associated
absorption systems towards \object{HE 0141--3932}  (Reimers at al. 2005).

\section{Conclusions}

We have analyzed  associated metal absorption systems
identified in the
UVES/VLT high resolution spectra ($FWHM \sim$6 \kms)
of four quasars
\object{J 2233--606}, \object{HE 1341--1020}, \object{HE 2347--4243},
and \object{Q 0329--385} with the aim to restore the spectral shape of the
underlying ionizing continuum in the EUV range $1 < E < 10$ Ryd. 
The systems were selected on base of criteria which
included the presence of lines of different ions and possibility to estimate
accurate covering factors and column densities. This ensured the reliability
and robustness of the restored SEDs. All systems are physically related to the  
quasar/host galaxy and, thus, the SEDs in question 
are representative for the quasar
outcoming radiation.  

The results obtained can be summarized as follows.
\begin{enumerate}
\item[1.] 
The ionizing spectra responsible for
the ionization structure in the analyzed systems
reveal a sharp intensity depression at $E > 4$ Ryd which is attributed to
the absorption in the \ion{He}{ii} Lyman continuum. 
The required column density of \ion{He}{ii} is of order $10^{18}$ \cm.
\item[2.] 
The most probable source of the \ion{He}{ii} opacity is the quasar 
accretion disk wind~-- a stratified flow
comprising dense near-equatorial gas accelerated by combined action of
radiation-pressure and magneto-centrifugal forces (seen as UV BAL
absorption) and more highly ionized shielding gas which protect the inner wind
regions from overionization by X-rays from the central source. All objects
considered here do not belong to BAL-quasars which means that the \ion{He}{ii}
opacity is due to the shielding gas. 
Assuming a pure power law SED at the wind base,
we can estimate the column density of neutral hydrogen corresponding to 
$N$(\ion{He}{ii})~$\sim$$10^{18}$ \cm: 
$N$(\ion{H}{i})~$\sim$ a few times $10^{16}$ \cm.
This amount of neutral hydrogen can be seen in a quasar spectrum as a
weak continuum depression starting at $\lambda 912$ \AA\ (rest-frame) 
and sometimes also as broad (stretching over 1000s \kms) 
and shallow absorption in \ion{H}{i} Ly-$\alpha$. 
In case of high metallicity (solar to oversolar) of the wind,
the absorption features due to resonance lines of \ion{O}{vi} 
in the FUV,  \ion{Ne}{vi}-\ion{Ne}{viii}
in the EUV, and \ion{O}{vii}-\ion{O}{viii} in the soft X-ray ranges 
are expected. 
In the UV spectra of quasars studied in the present paper,
we do find broad and shallow \ion{H}{i} Ly-$\alpha$  and 
\ion{O}{vi} $\lambda\lambda1031, 1037$ absorption features 
(\ion{Ne}{vi}-\ion{Ne}{viii} beyond the available wavelength coverage) 
as well as continuum depression at $\lambda \sim$912 \AA\ 
in the flux-calibrated spectra of \object{J 2233-606}
and \object{HE 2347--4243} 
(flux-calibrated spectra of other analyzed quasars are not available).
\item[3.]
In order to estimate the fraction of quasar outcoming radiation affected by
the \ion{He}{ii} Lyman continuum absorption, a systematic study of the predicted
continuum depression in the flux-calibrated high-resolution QSO spectra 
at wavelengths shorter than 912 \AA\ (rest-frame) should be
performed. This study is still to be done. However, observational data available
now both in UV and X-ray regions suggest 
that at least $\sim$50\% of the quasar radiation
passes through the gas opaque in the \ion{He}{ii} Lyman continuum. 
This means, that the outcoming power law
ionizing continuum has a pronounced intensity break at $E > 4$ Ryd with 
a depth of this break 
depending on the angle to the rotational axis of the accretion disk 
(the closer to the disk the deeper the break). 
This can influence the rate of the \ion{He}{ii} reionization 
in the intergalactic medium and partly explain inhomogeneous (patchy) 
ionization structure of the intergalactic \ion{He}{ii} observed 
at $z \sim$3. 
\item[4.]
Due to careful selection and detailed analysis of the absorption systems we
obtained quite accurate estimates of the elemental abundances in 
the absorbing gas. It is shown that the relative abundance of nitrogen 
to carbon, N/C, differs for the circumnuclear and interstellar gas 
of the host galaxy: at the similar oversolar metallicity
the circumnuclear gas is overabundant in nitrogen, whereas the galactic gas is
underabundant. This supposes different metal enrichment mechanisms which may be
due to different IMF in the regions near the center and 
at the periphery of a quasar host galaxy.
\end{enumerate}

 \begin{acknowledgements}
We would like to thank D. I. Nagirner, S. I. Grachev, and W. H. Kegel
for useful discussions.
I.I.A. and S.A.L. gratefully acknowledge the hospitality 
of the Shanghai Astronomical Observatory
and Hamburger Sternwarte while visiting there. This research has been  
supported by the RFBR grant No. 06-02-16489, 
and by the Federal Agency for Science and Innovations grant
NSh 9879.2006.2.
J.L.H. is supported by NFC of China No. 10573028, and by 973
program with No. 2007CB815402.
\end{acknowledgements}

\end{document}